\documentclass[10.5pt,a4paper]{amsart}

\makeatletter
\renewcommand{\email}[2][]{%
	\ifx\emails\@empty\relax\else{\g@addto@macro\emails{,\space}}\fi%
	\@ifnotempty{#1}{\g@addto@macro\emails{\textrm{(#1)}\space}}%
	\g@addto@macro\emails{#2}%
}
\makeatother

\usepackage{etoolbox}
\patchcmd{\section}{\scshape}{\bfseries}{}{}
\makeatletter
\renewcommand{\@secnumfont}{\bfseries}
\makeatother
\usepackage[margin=0.1cm]{caption}
\newenvironment{nouppercase}{%
  \renewcommand{\uppercasenonmath}[1]{}}{}
\usepackage{graphicx}  % needed for figures
\usepackage{dcolumn}   % needed for some tables
\usepackage{bm}        % for math
\usepackage{amssymb}   % for math
\usepackage{mathrsfs}
\usepackage{graphicx}
\usepackage{epstopdf}
\usepackage{color}
\usepackage{soul}
\usepackage{bm}
\usepackage{placeins}
\usepackage{amsmath}
\usepackage{mathtools}
\usepackage{float}

\def\i{{\rm i}}

\usepackage{color}

\def\p{\partial}
\def\be{\begin{eqnarray}}
	\def\ee{\end{eqnarray}}

\def\lp{\left(}
\def\rp{\right)}
\def\lb{\left[}
\def\rb{\right]}
\def\lcb{\left\{}
\def\rcb{\right\}}

\def\lap{\nabla^2}

\def\befi{\begin{figure}}
	\def\eefi{\end{figure}}

\def\no{\noindent}

\def\bce{\begin{center}}
	\def\ece{\end{center}}

\def\ba#1\ea{\begin{align}#1\end{align}}

\def\bsa#1\esa{\begin{subequations}\begin{align}#1\end{align} \end{subequations}}

\newcommand\rvec{\bm{r}}
\newcommand\evec{\bm{e}}

\newcommand\xvec{\bm{x}}

\newcommand\uvec{\bm{u}}

\newcommand\omegav{\boldsymbol{\omega}}

\usepackage{titlesec}

\titleformat*{\section}{\Large\bfseries}
\titleformat*{\subsection}{\large}
\usepackage{titlesec}
\titlelabel{\thetitle.\quad}

\usepackage{titlesec}

\titlespacing\section{0pt}{25pt plus 4pt minus 2pt}{14pt plus 10pt minus 10pt}
\titlespacing\subsection{0pt}{25pt plus 4pt minus 2pt}{44pt plus 10pt minus 10pt}
\titlespacing\subsubsection{0pt}{12pt plus 4pt minus 2pt}{44pt plus 2pt minus 2pt}

\graphicspath{ {Figures/} }
\usepackage{tikz}
\usepackage{amssymb,graphicx} 
\usepackage{amsaddr}
\usepackage{mathptmx} 
\usepackage{epstopdf}

\usepackage[numbers,sort,compress]{natbib}
  \def\ba#1\ea{\begin{align}#1\end{align}}

\DeclareMathAlphabet{\mathpzc}{OT1}{pzc}{m}{it}

\usepackage{mathptmx}
\DeclareMathAlphabet{\mathnew}{OMS}{cmsy}{m}{n}
\usepackage{mathtools} % for \text{}
\usepackage{mathrsfs}
\usepackage{graphicx}
\usepackage{epstopdf}
\usepackage{color}
\usepackage{soul}

\definecolor{darkblue}{RGB}{83,0,93}

\newsavebox{\astrutbox}
\sbox{\astrutbox}{\rule[-5pt]{0pt}{20pt}}

\def\lp{\left(}
\def\rp{\right)}
\def\lb{\left[}
\def\rb{\right]}
\def\lcb{\left\{}
\def\rcb{\right\}}

\def\ba#1\ea{\begin{align}#1\end{align}}

\def\bsa#1\esa{\begin{subequations}
		\begin{align}#1\end{align} \end{subequations}}

\usepackage{graphicx}% Include figure files
\usepackage{dcolumn}% Align table columns on decimal point
\usepackage{bm}% bold math
\date{\today}% It is always \today, today,
\usepackage[utf8]{inputenc}
\usepackage{textgreek}

\usepackage{mathtools}

\DeclarePairedDelimiter\floor{\lfloor}{\rfloor}

\numberwithin{equation}{section}
\usepackage[margin=1in]{geometry}
\usepackage{setspace}
\begin{document}
\setstretch{1.15}

\title[]{\LARGE{Active Cloaking in Stokes Flows via Reinforcement Learning}}

\author{Mehdi Mirzakhanloo $^{1}$, \ \ \ Soheil Esmaeilzadeh $^{2}$, \ \ \ and \ \ Mohammad-Reza Alam $^{1}$}
\address{$^{1}$Department of Mechanical Engineering, University of California, Berkeley, California 94720, USA\\ \vspace{5pt} $^{2}$Department of Energy Resources Engineering, Stanford University, California 94305, USA}
\curraddr{}
\email{reza.alam@berkeley.edu}
\thanks{}
\keywords{}
\date{}

% \author{Mehdi Mirzakhanloo}
% \address[]{Department of Mechanical Engineering, University of California, Berkeley, California 94720, USA}

% \author{Soheil Esmaeilzadeh}
% \address[]{Department of Energy Resources Engineering, Stanford University, CA, 94305, USA}

% \author{Mohammad-Reza Alam}
% \address[s]{Department of Mechanical Engineering, University of California, Berkeley, California 94720, USA}
% \email[s]{reza.alam@berkeley.edu}

\thanks{}
\keywords{}
\date{}

%\vspace{-15mm}

\begin{nouppercase}
\maketitle
\end{nouppercase}
%%%%%%%%%%%%%%%%%%%%%%%%%%%%%%%%%%%%%%%%%%%%%%%%%%%%%%%%%%%%%%%%%%%%%%%%%%%%%%

\begin{abstract}
\normalsize{Hydrodynamic signatures at the Stokes regime, pertinent to motility of micro-swimmers, have a long-range nature. This implies that movements of an object in such a viscosity-dominated regime, can be felt tens of body-lengths away and significantly alter dynamics of the surrounding environment. Here, we devise a systematic methodology to actively cloak swimming objects within any arbitrarily crowded suspension of micro-swimmers. Specifically, our approach is to conceal the target swimmer throughout its motion using cooperative flocks of swimming agents equipped with adaptive decision-making intelligence. Through a reinforcement learning algorithm, the cloaking agents experientially learn optimal adaptive behavioral policy in the presence of flow-mediated interactions. This artificial intelligence enables them to dynamically adjust their swimming actions, so as to optimally form and robustly retain any desired arrangement around the moving object without disturbing it from its original path. Therefore, the presented active cloaking approach not only is \textit{robust} against disturbances, but also is \textit{non-invasive} to motion of the cloaked object. We then further generalize the proposed approach and demonstrate how our cloaking agents can be readily used, in any region of interest, to realize hydrodynamic invisibility cloaks around any number of arbitrary intruders.}

\vspace{3mm}
\no 
\textbf{Key words:} \\ Cloaking, Swimming micro-Robots, Artificial Intelligence, Reinforcement Learning 

\end{abstract}

%\vspace{-20pt}
%\begin{tikzpicture}[remember picture, overlay]
%      \node[minimum width=12in,font=\large] at ([yshift=-1cm]current page.north)  {\small{On 16 July 2020, this paper was accepted for publication in the Journal of Fluid Mechanics.}};
%    \end{tikzpicture}

%%%%%%%%%%%%%%%%%%%%%%%%%%%%%%%%%%%%%%%%%%%%%%%%%%%%%%%%%%%%%%%%%%%%%%%%%%%%%%%%%%%%%%%%%%%%%%%%%%%%%%%%%%%%
\section{Introduction}

Living organisms in aquatic environments highly depend on detecting the fluid-mechanical signals caused by motions in the surrounding fluid \cite{pecseli2016plankton}. A broad range of swimming organisms, for instance, possess intricate sensors to directly measure the magnitude of disturbing flows (i.e. flow signatures) induced by nearby swimming objects \cite[see][]{yen1992mechanoreception, fields2002fluid, pecseli2016plankton}. This invaluable information is then used, as a `tool’, to detect the presence of nearby predators (preys), estimate their relative distance/size, and subsequently trigger an appropriate escape (catch) behavior \cite[see][]{bullock2008sensory, yen2015sensory, tuttle2019going}. For micro-swimmers, however, it was recently revealed \cite{mirzakhanloo2020stealthy} that an interacting flock can significantly suppress the induced disturbances (by $\sim 50\%$) once swarming in specific arrangements (referred to as \textit{concealed} modes). This finding inspires an even more intriguing question, that is whether a micro-swimmer can be actively cloaked, so as to remain undetectable (generating no trace) when passing through a host medium? Equivalently, is there a way to actively cloak random intruders to protect a sensitive region from disturbing effects of their induced fluid flows?

The concept of cloaking has long been of great interest to physicists, and remarkable progress has been made toward cloaking an object in the realm of electromagnetic waves \cite[see e.g.][]{pendry2006controlling,schurig2006metamaterial}, gravity waves \cite[see e.g.][]{alam2012broadband}, fluid flows \cite[see e.g.][]{urzhumov2011fluid, park2019hydrodynamic}, acoustics \cite[see e.g.][]{cummer2008scattering,zhang2011broadband}, quantum mechanics \cite[see e.g.][]{zhang2008cloaking}, thermodynamics \cite[see e.g.][]{guenneau2012transformation}, solid mechanics \cite[see e.g.][]{buckmann2015mechanical}, and even time \cite[see e.g.][]{boyd2012optical}.

Realizing a hydrodynamic invisibility cloak for a motile micro-swimmer, however, faces two fundamental challenges: (i) the \textit{dynamic} nature of the subject, that is continuously moving in arbitrary directions; and (ii) the long-range nature of underlying flow-mediated interactions, which can easily induce chaos and bring disorder to any potentially designed cloaking system that is initially set to a perfect order. The latter is further exacerbated for the case of cloaking specific subjects within a crowded suspension of micro-swimmers. In fact, the motion of each swimming object in the Stokes regime, perturbs the net flow field at the position of all nearby swimmers, and thus alters dynamics of the entire system. Therefore, to properly conceal swimming objects in a viscous environment, the cloak formation has to dynamically adjust in response to such non-linearly varying hydrodynamic loads. This leads us to the concept of \textit{active} cloaking.

Here we present the first demonstration of active cloaking in Stokes flows using cooperative flocks of micro-swimmers -- hereafter referred to as the `\textit{cloaking agents}'. In the presented method, hydrodynamic signature of the cloaked object -- hereafter referred to as the `\textit{intruder}', is actively suppressed (throughout its motion) by a group of cloaking agents that form specific arrangements around it. To optimally form (and robustly retain) such desired cloaking arrangements around a moving intruder, the agents indeed need to dynamically adjust their swimming actions. Therefore, we first provide a rigorous approach to systematically train the cloaking agents through a reinforcement learning algorithm. This experiential learning process, equips the agents with an optimal adaptive behavioral policy in the presence of flow-mediated interactions. The sequence of actions taken by the agents based on such an artificial intelligence, enables them to keep any arbitrary intruder concealed throughout its motion without disturbing it from its original path. Therefore, our active cloaking approach is also \textit{non-invasive} to the intruder's motion. We then further generalize our approach, and demonstrate how our cloaking agents can be readily used to realize hydrodynamic invisibility cloaks around any number of swimming objects, within any arbitrarily crowded suspension of micro-swimmers.

%%%%%%%%%%%%%%%%%%%%%%%%%%%%%%%%%%%%%%%%%%%%%%%%%%%%%%%%%%%%%%%%%%%%%%%%%%%%%%%%%%%%%%%%%%%%%%%%%%%%%%%%%%%%
\section{Interaction Dynamics in a Viscous Environment} \label{sec.Interactions}

\begin{figure}
	\centering 
	\includegraphics[width=0.72\textwidth]{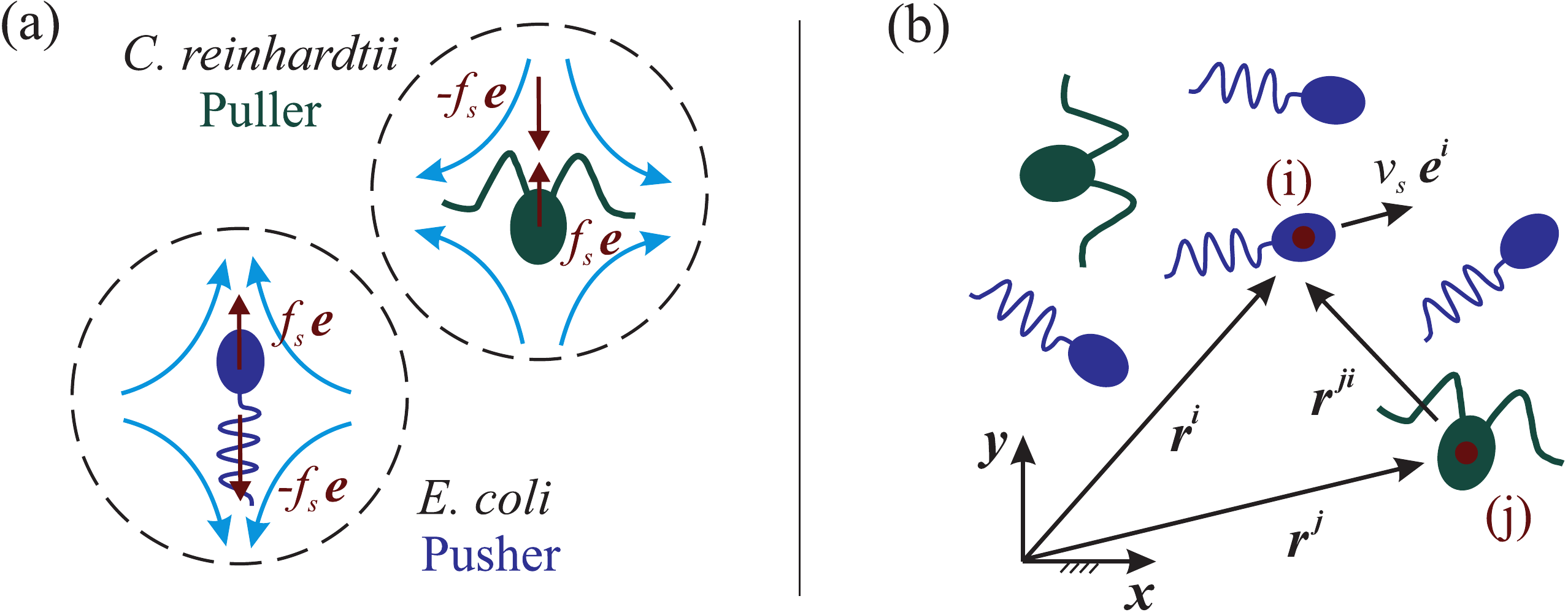}
	\caption {(a) Schematic representation of the archetypal puller and pusher swimmers. The force dipole ($\pm f_0 \bm{e}$) exerted by each swimmer to the surrounding fluid is shown by red arrows. Vector $\bm{e}$ denotes the swimming direction, and curly blue arrows demonstrate direction of the induced disturbing flows in each case. (b) Schematic representation for a system of interacting micro-swimmers. The position vector of swimmer $i \in \lcb 1 \dots N \rcb$, swimming with speed $V_s$ in direction $\bm{e^i}$, is denoted by $\bm{r}^i$ with respect to the fixed frame of reference. Here $N=6$, and the system includes both pusher and puller swimmers. }
	\label{figM0}
\end{figure}

Dynamics of the flow induced by a micro-swimmer (say e.g. a swimming micro-robot or a motile microorganism) can be described by the Stokes equations:
\ba\label{eq1}
\nabla P = \mu \ \lap \uvec + \bm{F}, \quad \nabla \cdot \uvec = 0.
\ea
where $\mu$ is dynamic viscosity of the surrounding fluid, $P$ represents the pressure field, $\uvec$ is the velocity field, and $\bm{F}$ is the notion of body forces per unit volume. 
In the absence of external (magnetic or electric) fields, a self-propelled buoyant micro-swimmer exerts no net force/torque to the surrounding fluid. 
Therefore, in the most general form, far-field of the disturbing flow induced by a micro-swimmer is well-described by the flow of a force dipole ($\pm f_s \evec$) composed of the thrust force (generated by the swimmer's propulsion mechanism) and the viscous drag acting on its body (Fig. \ref{figM0}a). 
To be more precise, for a micro-swimmer swimming with speed $v_s$ toward direction $\bm{e}$ through an unbounded fluid domain, the induced flow field can be formulated \cite[see e.g.][]{elgeti2015physics} as
$\uvec_{\text{SD}} = \mathscr{D}/(8\pi \mu \ r^3) \lb -1+ 3 \lp \rvec \cdot \bm{e}/r \rp^2 \rb \rvec$,
where $\rvec = \xvec- \xvec_0$ for any generic point $\xvec$ in space, and $\xvec_0$ represents the instantaneous position of the swimmer. The dipole strength, $|\mathscr{D}| \sim v_s L_s^2 \sim f_s L_s$, has a positive (negative) sign for pusher (puller) swimmers, and its value can be inferred from experimental measurements \cite[see e.g.][]{drescher2011fluid}. Here the characteristic length $L_s$ is on the order of swimmer dimensions, and we use $\lcb U_s = v_s, L_s, T_s = L_s/v_s \rcb$ to make the quantities dimensionless throughout this work.

Swimming in the Stokes regime is significantly affected by the presence of long-range hydrodynamic interactions. In fact, disturbing flows induced by the motion of each micro-swimmer highly affects dynamics of other nearby swimming objects. Here we take into account both \textit{hydrodynamic} and \textit{steric} interactions between the swimmers. 
To this end, let us consider a generic flock of `$N$' interacting micro-swimmers as presented schematically in Fig. \ref{figM0}(b). 
For any swimmer $i \in \lcb 1, \dots,N \rcb$ in the flock, dynamics of the position vector, $\rvec^i$, and the swimming direction, $\evec^i$, are governed by
\begin{subequations}\label{eq2}
	\begin{align}
		\dot{\rvec}^i &= v_s \evec^i + \sum_{j\neq i} \lb \uvec^{j} \lp \rvec^i \rp + \frac{1}{6\pi \mu L_s} \ \bm{\mathcal{F}}^{j} \lp \rvec^i \rp  \rb + \uvec_{f} \lp \rvec^i \rp , \label{eq2a}\\
		\label{eq2b}
		\dot{\evec}^i &= \lcb \frac{1}{2} \ \omegav_f\lp \rvec^i \rp + \frac{1}{2} \sum_{j\neq i} \omegav^{j}\lp \rvec^i \rp + 
		\mathcal{B} \evec^i \times  \lb \mathtt{E}_f\lp \rvec^i \rp + \sum_{j\neq i} \mathtt{E}^{j}\lp \rvec^i \rp \rb \cdot \evec^i \rcb \times \evec^i,
	\end{align}
\end{subequations}
where $\uvec^{j} \lp \rvec^i \rp = \uvec_{\text{SD}} \lp \rvec^i -\rvec^j \rp $ is the disturbing flow induced by swimmer `j' at the position of swimmer `i', and $\omegav^j = \nabla \times \uvec^j$ is the corresponding vorticity field. 
Similarly, $\uvec_{f} \lp \rvec^i \rp$ denotes the background (external) flow at the position of swimmer `i', and $\omegav_f = \nabla \times \uvec_f$ is the corresponding vorticity field.
Here each swimmer is modeled as an infinitesimal spheroid of aspect ratio $\lambda$, for which the Bretherton constant \cite[see e.g.][]{kim2013microhydrodynamics} is defined as $\mathcal{B} = \lp \lambda^2-1 \rp / \lp\lambda^2+1\rp$.
The rate of strain tensor, defined as $\mathtt{E} \lp \rvec^i \rp = (1/2)\lb \nabla \uvec (\rvec^i) + \nabla \uvec^\top (\rvec^i) \rb$, is also denoted by $\mathtt{E}^{j}$ and $\mathtt{E}_f$, for $\uvec^{j}$ and $\uvec_{f}$, respectively.
To regularize our dipole models (diverging as $1/r^2$), we also define a purely repulsive force \cite[c.f.][]{ryan2011viscosity}, based on Lenard-Jones-type potential:
\ba \label{eq3}
\bm{\mathcal{F}}^{j} \lp \rvec^i \rp = -\p \mathscr{L}^{ji} /\p \rvec^{ji}, \quad
\mathscr{L}^{ji} = 4 \epsilon_{L} \lb \lp \frac{\sigma}{r^{ji}} \rp ^{12} - \lp \frac{\sigma}{r^{ji}} \rp ^{6} \rb + \epsilon_{L},
\ea 
where $r^{ji} = |\rvec^i-\rvec^j|$, the constant $\sigma$ indirectly specifies the equilibrium distance (i.e. $2^{1/6} \sigma \sim 2L_s$), and $\epsilon_{L}$ tunes strength of the steric interactions. This ensures the excluded volume constraints, and thus pushes the swimmers away from each other when approaching closer than a certain distance (determined by $2^{1/6} \sigma$).

It is worth noting that the presented framework \eqref{eq2} can be readily extended by incorporating a more detailed description of $\uvec^{j} \lp \rvec^i \rp$.
In particular, near-field of the flow induced by micro-swimmers can be described more accurately through including an appropriately chosen combination of higher order terms from the multipole expansion \cite[see e.g.][]{spagnolie2012hydrodynamics,ghose2014irreducible}.
However, in this work, we are interested in the \textit{span} of swimmers' induced disturbances and their consequent detection region, for which far-field of the flow is of primary interest.
Therefore, as discussed earlier, the disturbing flow induced by each micro-swimmer is formulated here, in the most general form, as the flow of a force dipole.
This simple model has been validated and widely used in the literature \cite[see][]{lauga2009hydrodynamics,elgeti2015physics}.
In particular, for archetypal pusher and puller swimmers (i.e. E. coli bacteria and C. reinhardtii alga) used as benchmarks throughout this work, the validity of this model is further confirmed via direct comparison to the disturbing flows (experimentally) measured around individual swimming cells \cite{drescher2010direct,drescher2011fluid}.
In the case of E. coli bacteria, for instance, values of $f_s=0.42$ pN and $L_s=1.9$ $\mu$m, have been suggested \cite{drescher2011fluid} in agreement with resistive force theory \cite{darnton2007torque} and optical trap measurements \cite{chattopadhyay2006swimming}. 
The corresponding dipole model (with $\mathscr{D}\sim f_s L_s$) is proven to be highly accurate in predicting magnitude of the induced disturbing flows, even at distances comparable to characteristic length of the swimmer \cite{drescher2011fluid}. 
To be more specific, the predicted values are in agreement with those experimentally measured around an individual swimming cell at ranges $r\gtrsim 3L_s$.
At shorter distances, the induced disturbing flows are shown to be less significant than those estimated by the dipole model \cite{drescher2011fluid}, and thus steric interactions (`soft' collisions) are known to dominate the hydrodynamic interaction. 
Therefore, it is expected that the dipolar hydrodynamics, when supplemented by Lennard-Jones-type excluded volume interactions \eqref{eq3}, well-describe interaction dynamics in a semi-dilute suspension (see e.g. \cite{ryan2011viscosity,ryan2013kinetic}).
This agent-based model has also been reported \cite{ariel2018collective} successful in predicting collective dynamics of two-dimensional bacterial swarms, for swimmer concentrations (in terms of surface fractions) of up to 0.6.

To keep the model simple and tractable in this work, we specifically consider swimmers with spherical body shapes (with $\mathcal{B}=0$) for our swimmers. 
Nevertheless, the presented framework \eqref{eq2} explicitly incorporates the effects of swimmers' geometry. Thus, it can be used to describe the interaction dynamics of micro-swimmers with different shapes.
In the case of ellipsoidal body shapes with eccentricity $\lambda$, for instance, the Bretherton constant can be evaluated as $\mathcal{B} = \lp \lambda^2-1 \rp / \lp\lambda^2+1\rp$. 
This stands for the values of $\mathcal{B} =$ 0/1 for spheres/needles, and takes positive (negative) values for prolate (oblate) objects.

%%%%%%%%%%%%%%%%%%%%%%%%%%%%%%%%%%%%%%%%%%%%%%%%%%%%%%%%%%%%%%%%%%%%%%%%%%%%%%%%%%%%%%%%%%%%%%%%%%%%%%%%%%%%
\section{Cloaking a micro-Swimmer in Stokes Flows}  \label{sec.Cloaking}

A cloak is commonly referred to as a patch enclosing an object to make it invisible. 
In an aquatic environment, however, swimming organisms nearly always use fluid disturbances (caused by motions in the surrounding fluid), as hydrodynamic signals to detect the nearby objects.
Therefore, to be invisible in such environments, one needs to generate no disturbances to the ambient fluid.
Accordingly, here we define the cloak as a patch virtually covering/enclosing a swimmer to cancel out its induced disturbances. This will then keep the enclosed swimmer `invisible' to others in the medium -- by stifling its flow signature. Similar to other forms of cloaking, a suitable cloak is also required to be omni-directional, that is, to keep itself and the enclosed swimmer `invisible' from any direction. 
It worths highlighting that the induced disturbing flows have a long-range nature at the Stokes regime pertinent to motility of micro-swimmers. 
This means that despite high Reynolds number motions in which hydrodynamic signature is relatively confined to the immediate neighborhood, the motion of an object at a low Reynolds number regime can be felt tens of body-lengths away. As a result, movements of micro-swimmers significantly alter their surrounding environment.
Therefore, specific to biological applications, the cloaking may be more important in \textit{protecting} sensitive regions (e.g. organs) against disturbing flows induced by intruders -- such as passing-by swimming organisms or deployed biomedical micro-robots. This, indeed, is equivalent to making such intruders hydrodynamically invisible (i.e. cloaked) within the surrounding environment.

\begin{figure}
	\centering 
	\includegraphics[width=0.8\textwidth]{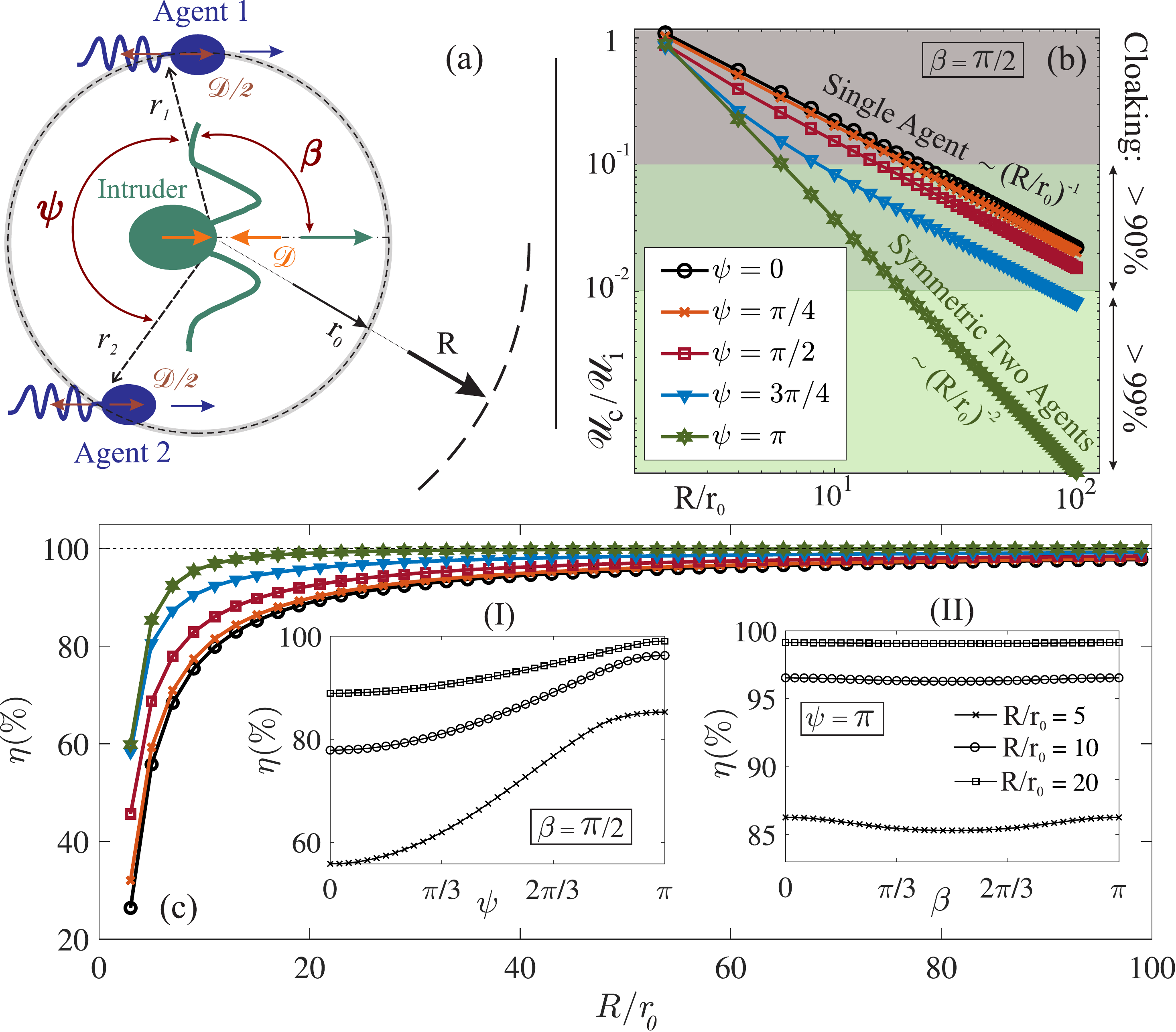}
	\caption {\textbf{(a)} Schematic representation of the presented method for cloaking an intruder (here a puller swimmer shown in green) using properly positioned single ($\psi=0$) or double ($\psi \neq 0$) cloaking agents (here the pusher swimmers shown in blue). To compute the cloaking efficiency of different arrangements, the induced fluid disturbances are measured over a surrounding (dashed) ring of radius $R$. The dipole strength of each smart agent is half of the intruder's (i.e. $\mathscr{D}/2$). The agents are positioned on a (highlighted dashed) ring of radius $r_0$ around the intruder -- i.e. their separation distance from the intruder is set to $r_0$, and their relative position vectors are denoted by $\bm{r_1}$ and $\bm{r_2}$, respectively. The angle between these two vectors is denoted by $\psi$ and represents a measure of symmetry. Also, $\beta$ is defined as the angle between intruder's swimming direction (green arrow) and the position vector $\bm{r_1}$.
		\textbf{(b)} Magnitude of the net fluid disturbances ($\mathscr{U}_c$) induced by the cloaked intruder and its cloaking agents, normalized by that of an isolated intruder ($\mathscr{U}_i$), is plotted for various cloak arrangements as a function of distance from the swimmer (R). Specifically, $\mathscr{U}_c/\mathscr{U}_i$ is plotted as a function of $R/r_0$ for $\psi=$ 0, $\pi/4$, $\pi/2$, $3\pi/4$, and $\pi$, where $\beta$ is set to $\pi/2$.
		\textbf{(c)} The cloaking efficiency of the arrangements associated with each of the $\psi$-angles presented in panel (b), is plotted as a function of $R/r_0$. Insets also demonstrate how the cloaking performance (in terms of $\eta \%$) varies with $\psi$ (I) and $\beta$ (II), respectively. For all cases presented in (I), $\beta$ is fixed to $\pi/2$, and for those presented in (II), $\psi$ is fixed to $\pi$. The presented values are measured at $R/r_0 =$ 5, 10, 20.}
	\label{figM1}
\end{figure}

We assess the effectiveness of an invisibility cloak in Stokes regime, by evaluating its efficiency in stifling the swimmer's induced disturbing flows. 
A measure of distortion caused by swimming objects to the ambient fluid can be obtained through directly computing the mean disturbing flow-magnitude ($\mathscr{U}$) over a surrounding ring of radius $R$ (c.f. Fig. \ref{figM1}a) -- here denoted by $\mathcal{C}(R)$. The cloaking efficiency ($\eta$) can then be calculated as
\ba \label{eq12}
\eta = 1- \frac{\mathscr{U}_c}{\mathscr{U}_i}, \quad \mathscr{U} = \frac{1}{2\pi R} \oint_{\mathcal{C}(R)} |\bar{\uvec}| \ \text{ds},
\ea	
where $\mathscr{U}_c$ measures the induced disturbances once the cloak is implemented (i.e. corresponding to the net flows generated by the swimmer and its implemented cloak), and is directly compared to disturbances induced by the isolated swimmer ($\mathscr{U}_i$) -- when no cloaking is in place. Here, `$\text{ds}$' is the differential length along the surrounding ring, and $\bar{\uvec}=\uvec/v_s$ represents the net dimensionless flow field induced by the swimmer/system.

The simplest strategy in cloaking a single swimmer (an \textit{intruder}) at any given instance of time is to properly position, next to it, another swimmer of the same dipolar strength, but from the opposite type. 
This basically replaces the dipolar (vanishing as $1/r^2$) leading order of the overall disturbing flows (in the far-field) with a quadrupole (rapidly decaying as $1/r^3$), and thus can effectively cloak the intruder by drastically stifling its induced disturbances in the ambient fluid. 
To give an example, a single cloaking agent positioned next to the intruder with a separation distance of $r_0$, dramatically reduces magnitude of the overall detectable disturbing flows (measured by $\mathscr{U}$) to 11.6\% and 4.4\% of its original value, once measured at distances $R/r_0=$ 20, and 50 from the intruder, respectively (Fig. \ref{figM1}-b for $\psi=0$). This is equivalent to more than 88.4\% and 95.6\% in cloaking efficiencies at the ranges of $R/r_0$ $\geq 20$, and $\geq 50$, respectively (Fig. \ref{figM1} for $\psi=0$). 
For this simple approach, however, the underlying hydrodynamic interaction between the intruder and its cloaking agent will inevitably disturb the intruder from its original path. Therefore, using a single cloaking agent, although effective, is not the smartest approach.

Here, we are particularly interested in implementing non-invasive active cloaking, where the cloaked swimmer (i.e. the intruder) is not deviated from its original path by the cloak implementation. This is fundamentally impossible using only a single cloaking agent.
Alternatively, a proper positioning of two \textit{half-sized} cloaking agents around an intruder (e.g. see Fig. \ref{figM1}-a for $\psi=\pi$ and $\beta=\pi/2$), can potentially prevent its deviation from the original path. This also significantly improves the cloaking efficiency (by switching the leading order in the far-field of induced disturbing flows to an octupole rapidly vanishing as $1/r^4$) compared to the single-agent method (with quadrupolar far-field decaying as $1/r^3$).

The schematic representations of cloaking an intruder using single ($\psi=0$) and double ($\psi \neq 0$) cloaking agents are presented in Fig. \ref{figM1}(a). 
We also present cloaking performances corresponding to various arrangements of the agents around an intruder in Fig. \ref{figM1} -- both in terms of the relative magnitude of induced disturbances ($\mathscr{U}_c/\mathscr{U}_i$) and the cloaking efficiency ($\eta$). 
Note that the presented efficiencies correspond to a set of optimally designed cloaking agents which can provide an ideally canceling set of dipole strengths for the specific intruder of interest.
Our results show that the effect of angular positioning (measured by $\beta$) on cloaking efficiency is negligible (Fig. \ref{figM1}c-II), whereas $\psi$ (as a measure of symmetry) primarily controls the efficiency (Fig. \ref{figM1}c-I). 
Variation of $\psi$ in the range of $\lb0,\pi \rb$ basically covers the whole spectrum from single-agent ($\psi = 0$) to symmetric double-agent ($\psi = \pi$) cloaking. 
It is worth highlighting that the symmetric arrangement of cloaking agents on the sides of an intruder (i.e. the arrangement with $\psi=\pi$ and $\beta=\pi/2$ in Fig. \ref{figM1}-a), represents not only a \textit{non-invasive}, but also the most \textit{efficient} double-agent cloaking strategy. The latter, in particular, is not immediately clear a priori, and has been revealed through the presented results (Fig. \ref{figM1}).
In practice, such an optimal arrangement can potentially be used to realize a virtually perfect \textit{non-invasive} cloak, the efficiency of which reaches beyond 99\%, and 99.9\% at distances $R/r_0$ $\geq$ $20$ and $50$, respectively (Fig. \ref{figM1}).

\textit{Active cloaking} of an arbitrarily moving intruder in the Stokes regime, however, still remains a challenge.
For naive (i.e. non-smart) pair of cloaking agents, even if they are initially placed accurately in their predefined positions, the arrangement will soon be distorted due to the presence of long-range hydrodynamic interactions -- causing the swimming agents to either diverge from one another or collide. 
In fact, disturbing flows induced by the intruder (although canceled out by its cloak in the far-field) are internally disruptive for the cloak itself -- i.e. disturbs the agents' arrangement and thus breaks the cloak. 
Therefore, to realize \textit{active} cloaking in the presence of such complex flow-mediated interactions, key questions still remain to be addressed in the following section:
First, what is the optimal sequence of actions (corresponding to the shortest non-invasive path) for each cloaking agent to position itself in a desired arrangement around an arbitrary intruder? 
Then, what is the optimal behavioral policy for the cloaking agents to dynamically adjust their swimming actions in response to non-linearly varying hydrodynamic loads, so as to robustly keep their concealing arrangements around an intruder? 
Lastly, to what extent is the presented approach generalizable? That is, can we use it to actively cloak multiple arbitrary intruders within crowded suspensions, where each swimmer moves toward an arbitrary direction and experiences frequent close encounters?

%%%%%%%%%%%%%%%%%%%%%%%%%%%%%%%%%%%%%%%%%%%%%%%%%%%%%%%%%%%%%%%%%%%%%%%%%%%%%%%%%%%%%%%%%%%%%%%%%%%%%%%%%%%%
\section{Learning to Cloak Random Intruders via Reinforcement Learning} \label{sec.Learning}

Here we present a systematic methodology to equip micro-swimmers with adaptive decision-making intelligence, in response to flow-mediated interactions. We then use these smart agents to elucidate \textit{active} cloaking of arbitrary intruders within a crowded environment. 
Note that once an agent is equipped with such an artificial intelligence, it will consider potential consequences of its actions when making any decision -- hence, is called a \textit{planning} agent.
One may alternatively suggest the implementation of an external active control to lead each of the employed swimming micro-robots (i.e. cloaking agents) toward the desired arrangements.
However, the efficacy of such approaches suffer from nonlinearly varying hydrodynamic loads due to the presence of flow-mediated interactions. 
In fact, the dynamical system representing a group of interacting micro-swimmers, is a complicated four-way-coupled system, where any tiny perturbation/deviation in prescribed motions of the swimmers can induce unpredicted complex choreographies \cite[see e.g.]{mirzakhanloo2018hydrodynamic}. 
It also worths noting that the \textit{reflex} agents (by definition) are incapable of identifying \textit{optimal} pathways to form non-invasive cloaks, especially in crowded suspensions or in the presence of obstacles.
Therefore, here our approach is to let the cloaking agents learn optimal action policies by their own experience through a reinforcement learning algorithm \cite{sutton2018reinforcement}.
By accumulating experience, our cloaking agents learn how to optimally collaborate, and dynamically adjust their swimming actions, to form a robust non-invasive cloak for any randomly moving intruder.

The implemented reinforcement learning algorithm was initially inspired by the concept of animal learning \cite{niv2006choice}, and has been shown effective in learning previously unknown strategies, solely based upon the received feedback on performance \cite{tesauro1995temporal}.
The great potential of this approach has also been recently demonstrated in fish schooling \cite{gazzola2016learning}, soaring of birds through turbulent environments \cite{reddy2016learning}, flow navigation of gravitactic particles \cite{colabrese2017flow}, and many other applications in the realm of active matters \cite{cichos2020machine,tsangroads2020,tsang2020self}.

\begin{figure}
	\centering 
	\includegraphics[width=0.9\textwidth]{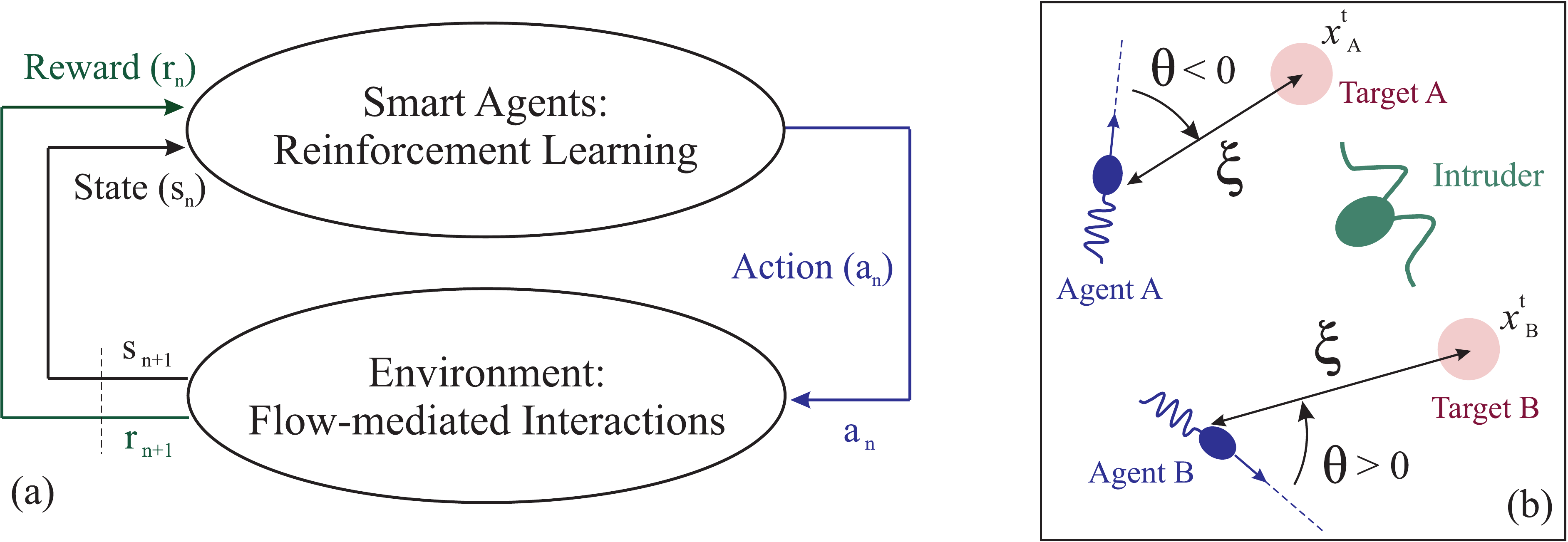}
	\caption {(a) Schematic representation of the Reinforcement Learning algorithm coupled with the presented flow-mediated interaction dynamics in viscous environments. (b) Schematic illustration of the state-space $(\xi,\theta)$, when two of our smart micro-swimmers (agents A and B shown in blue) attempt to form a desired active cloak around a randomly moving intruder (green).}
	\label{figM2}
\end{figure}

Here, we formulate the reinforcement learning algorithm as a Markov Decision Process (MDP).
This allows us to use the Q-learning framework, which not only benefits from algorithmic simplicity, but also is proven \cite{watkins1992q} to converge to an \textit{optimal} behavioral policy.
In this framework, the \textit{agent} (here a micro-swimmer capable of decision-making) gradually learns the optimal behavioral policy through exploring the environment (see Fig. \ref{figM2}a).
At any given instance of time (the $n$th learning step), the agent is able to sense some information about the environment (\textit{state}, $s_n$), depending on which, it will choose an \textit{action} ($a_n$) according to its current \textit{policy} ($\pi_n$). Taking this action will then transit the agent to a new state ($s_{n+1}$), and it will be given a \textit{reward} ($r_{n+1}$) quantifying its immediate success (Fig. \ref{figM2}a). 
The experience acquired by the agent (after going through each learning step) is stored as an action-value function $Q(s,a)$, in the Quality matrix `$Q$' -- hence the name of algorithm. 
To be more precise, this Q-matrix encodes action-value function under the current policy, which represents the expected sum of discounted future rewards when taking the action $a_n$ at the current state ($s_n$) and following the current policy ($\pi_n$) thereafter; i.e.,
\ba
Q \lp s_n,a_n \rp = r_{n+1} + \gamma r_{n+2} + \gamma^2 r_{n+3} + \dots \ . 
\ea
This Q-matrix is updated throughout the learning process, and is proven cite{watkins1992q} to encode \textit{optimal} action-value function ($Q^*$) at convergence.
The optimal behavioral policy ($\pi^*: s_n \rightarrow a_n$) is then readily available. To be more precise, in any given state ($s_n$), the optimal action ($a_n^*$) is the one that maximizes the expected sum of future rewards (which is encoded in $Q^*$). This policy ($\pi^*$), in fact, serves as the agent's adaptive decision-making intelligence.

Here we implement an off-policy control using the described Q-learning scheme. 
During the learning phase, an $\epsilon$-greedy behavioral policy has been employed to ensure exploration of new solutions while appropriately exploiting the gained knowledge. 
Specifically, at the $n^{\text{th}}$ learning step, the policy ($\pi_n: s_n \rightarrow a_n$) is to choose the action ($a_n$) that maximizes current evaluation of the action-value function ($Q$). Except for a small probability ($\epsilon$), in which case, a random action will be chosen (independent of $Q$) from the set of possible actions ($\mathcal{A}$) to further explore the state-space; i.e.
\ba
a_n =
\begin{cases}
	\text{arg} \max_{a} Q(s_n,a) & \text{probability of $1-\epsilon$}\\
	\text{random action $a \in \mathcal{A}$} & \text{probability of $\epsilon$}\\
\end{cases}      \,. 
\ea

After going through each learning step, the current estimation of the action-value function ($Q$) is then updated according to
\ba
Q \lp s_n,a_n \rp  \leftarrow Q \lp s_n,a_n \rp  + \alpha \lb r_{n+1} + \gamma \max_{a} Q \lp s_{n+1},a \rp - Q \lp s_n,a_n \rp  \rb,  
\ea
where the learning rate, $0 \leq \alpha \leq 1$, specifies the rate at which, previously learned experience is overwritten by the newly gained information. The discount factor, $0 \leq \gamma <1$, determines to what extent the value of expected future rewards is incorporated into the agent's decision-making:
for $\gamma \rightarrow 0$ ($\rightarrow 1$) its behavior will be myopic (farsighted) tending to maximize immediate (future) rewards.  
Here, the state-transition ($s_n \rightarrow s_{n+1}$), i.e. evolution of the system dynamics, is directly simulated taking into account the four-way coupled nature of the system through flow-mediated interactions (see section \ref{sec.Interactions} for details). 
Once in the new state ($s_{n+1}$), the agent will then take another action, based on the updated policy, and the process will be repeated (Fig. \ref{figM2}a).
Note that the updated action policy ($\pi_{n+1}$) in the new state will still be the same $\epsilon$-greedy policy, yet based on the updated estimation of $Q$. Therefore, the behavioral policy is continuously improved through the learning phase, and eventually converges to the optimal behavioral policy ($n \rightarrow \infty: \ \pi_n \rightarrow \pi^*$). 
We then employ a purely deterministic (greedy) behavioral policy (i.e. $\epsilon=0$) for our cloaking agents, according to the obtained optimal action-value function, $Q^*$, to further assure the success of any assigned task.

Let us now consider the problem of cloaking arbitrarily moving intruders using pairs of swimming agents. We first outline a rigorous approach to equip our cloaking agents with adaptive decision-making intelligence -- i.e. realizing the concept of \textit{smart} micro-swimmers.
As a benchmark, we consider puller intruders (say e.g. motile C. reinhardtii cells) swimming in an infinite two-dimensional fluid domain. Nevertheless, the results will be similar for pushers, and our study can be inherently generalized to a 3D domain.
Each of the cloaking agents is modeled as a self-propelled micro-swimmer of the opposite type (here pushers) with a half-sized dipole strength ($\mathscr{D}/2$). 
We are particularly interested in realizing the \textit{active} (and robust) form of the most effective cloak presented in section \ref{sec.Cloaking} (Fig. \ref{figM1}) -- for which, $\eta>99\%$ at $R/r_0 \geq 20$. Therefore, the approach is to position two cloaking agents symmetrically arranged (with a predefined separation distance $r_0$) on the sides of each intruder throughout its motion. 
This non-invasive cloak will effectively keep the enclosed swimmer hydrodynamically `invisible' within the surrounding fluid.

In order to form this active cloak, our cloaking agents must learn optimal behavioral policy in the presence of long-range hydrodynamic interactions. 
A pair of these smart micro-swimmers can then be assigned to cloak each of the randomly moving swimmers within any arbitrarily crowded suspension. 
The learning objective for each of these two cloaking agents can be translated into two specific tasks:
(i) learning how to optimally catch the assigned (randomly moving) intruder by forming a desired cloaking arrangement around it;
and once reached to the target positions (e.g. points A and B in Fig. \ref{figM2}b), then (ii) learning how to keep on following the assigned intruder while robustly retaining the desired arrangement. 
Hereafter, we refer to these two consecutive learning tasks as to `\textit{catch}' and `\textit{follow}' a random intruder -- both required in realizing an active cloak. 
Note that optimal execution of the former requires the cloaking agents to identify shortest non-invasive paths toward positioning themselves (from anywhere in the state space) into predefined desired arrangements around any moving subject.

To accomplish these tasks, each swimming agent can sense (or is provided with) an estimation of its distance and relative orientation with respect to the intruder. 
In practice, a range of imaging techniques \cite[see][]{pane2019imaging} can be used to feed such information into the agent's decision-making unit.
For instance, magnetic resonance imaging (MRI) has successfully been integrated into robotic platforms to provide the information required for real-time navigation of untethered micro-robots (or bacteria) swimming through human micro-vasculature \cite[see e.g.][]{martel2009mri,martel2009flagellated}.

The acquired information by each agent, mathematically translates to the normalized distance $\xi = |\xvec^t-\xvec|/L_s$ and relative orientation $\theta$ with respect to the assigned target point, as depicted in Fig. \ref{figM2}(b) for the catching phase.
This representation of the state-space is also consistent with previous studies (e.g. \cite{gazzola2016learning}) on swarm control of swimming agents.
Note that once the agents successfully \textit{catch} the intruder (by forming the desired cloaking arrangement around it), they still will need to \textit{follow} the intruder, so as to actively keep the desired arrangement throughout its motion. Therefore, once in the cloaking positions, $\theta$ is defined as the deviation angle that each agent senses for its own swimming direction relative to the assigned intruder's.

Although the micro-swimmers are free to move in a continuous two-dimensional space, here we map the state-space, $s = (\xi, \theta)$, on a finite-size 2D pseudo-grid. 
This way of representing the state-space dramatically mitigates the curse of dimensionality, and makes the learning process computationally feasible.
Also, from the practical point of view, the agents will then only need an estimation of their distance and deviation from their assigned target points. The required accuracy of such estimation is specified by the grid-size.
Therefore, let us consider the discretized representation of our state-space ($\mathcal{S}$), which is the product of two finite-size subsets, i.e. $\mathcal{S} = \mathcal{S}_{\xi} \times \mathcal{S}_{\theta}$. Specifically, $\mathcal{S}_{\xi}$ represents the distance $\xi$ as $n_{\xi}$ discrete states within the range of $\Delta \xi$, and $\mathcal{S}_{\theta}$ indexes the angle $\theta$ as $n_{\theta}$ discrete states within the range of $\Delta \theta$.
Hence, at every instance of time ($t_n$), the state of an agent is represented by a tuple $s_n = (\xi_n, \theta_n)$, where 
\bsa
\xi_n = \min \ ( \Delta \xi, \ \floor*{ \xi n_{\xi}/\Delta \xi } \cdot \Delta \xi/n_{\xi}  ), \\
\theta_n = \min \ ( \Delta \theta, \ \floor*{ \theta n_{\theta}/\Delta \theta } \cdot \Delta \theta/n_{\theta}  ).
\esa
For instance, here we set $n_{\xi}=100$, $\Delta \xi = 50$, $n_{\theta}=36$, and $\Delta \theta = 2\pi$. This means a total of 3,600 discrete possible states for each agent.

The deployed cloaking agents are able to swim in three distinct speeds: $v^0$ (nominal), $v^+ = v^0 + \delta v$ (fast), and $v^- = v^0 - \delta v$ (slow). They can also instantly turn to right ($\theta \leftarrow \theta + \delta \theta$) or left ($\theta \leftarrow \theta - \delta \theta$) with different choices of angle $\delta \theta$.
Therefore, the motion of each swimming agent during the learning process can be described as a repeated occurrence of the following two-step process:
(i) the agent updates its swimming velocity (through either a change in orientation by $\delta \theta$, or a modification in speed by $\delta v$) according to the current policy; and then
(ii) it moves with the new velocity for a fixed time $\tau_r$ (=1 in non-dimensional time units scaled by $T_s = L_s/v^0$).
Note that each agent takes an \textit{exclusive} swimming action (e.g. turning to the right/left with an angle $\delta \theta$, or speeding up/down with $\delta v$) based on the current policy ($\pi_n$), at intervals of $\tau_r$ during the learning process. Once an action is taken, the agent swims steadily forward (i.e. `runs'), until a new action (potentially a change in orientation, i.e. `tumbling') is taken after $\tau_r$.
This is inspired by the observed behavior of natural swimming microorganisms, and is reminiscent of the so-called \textit{run-and-tumble} locomotion.
In practice, realization of the smart version of this locomotion dynamics seems also feasible in the context of artificial and model micro-swimmers. The recently proposed Quadroar swimmer \cite{jalali2014versatile, mirzakhanloo2018flow}, for instance, propels on straight lines (runs), and can perform full 3D reorientation (tumbling) maneuvers \cite{saadat2019experimental}.

Here, the nominal swimming speed for the agents is set to that of the intruders (i.e. $v^0 = v_s$), and we consider $\delta v = 0.05 \ v^0$ (0.5 $v^0$) while following (catching) an intruder. 
Nevertheless, our numerical experiments reveal that a learned policy (under this specification) can be readily adopted by agents with different speed characteristics ($v^0$, $\delta v$), and so in cloaking intruders with various swimming speeds (see Appendix A.1 for details).
Furthermore, the space of possible values for $\delta \theta$, inherently covers the entire range of $\lb 0, \pi\rb$.
However, selecting only a finite-size set of options for $\delta \theta$, substantially reduces the computational costs associated with the recurrent identification of optimal actions, during the reinforcement learning process.
Therefore, here we assume the agents are only able to turn right/left with $\delta \theta \in \lcb \pi/18,\pi/4,\pi/2 \rcb $. 
This finite-size simplified set is particularly devised to provide each agent with the ability to:
(i) make major changes in the swimming direction (with $\delta \theta = \pi/2$), typically done in early stages of a catching process;
(ii) perform moderate re-orientations (with $\delta \theta = \pi/4$) commonly used during a catching process; and
(iii) fine tune slight deviations from a desired orientation (with $\delta \theta = \pi/18$), which is particularly required at final stages of a \textit{catching} process or while \textit{following} an intruder.
Therefore, our set of possible actions ($\mathcal{A}$) consists of nine specified swimming actions. This, along with the described discretized state-space, result in 32,400 entries for the space of state-action pairs, to be used in evaluating the action-value function $Q(s_n,a_n)$.
We note that the frequency of taking an action by each agent (i.e. $\tau_r^{-1}$) can also be tuned according to the system of interest.

For the swimming agents, realization of an active cloak around an arbitrary intruder is translated into taking an optimal sequence of actions to actively catch and follow the assigned target points in a predefined arrangement. 
The immediate success of each agent in satisfying this assigned task is mathematically interpreted as a numerical reward signal 
\ba \label{eq4.4}
r_n = [1/(\xi_n+\delta \xi) - \xi_n] + [\text{\textdelta}(v_n-v^0)-1] + \mathscr{C}_n + \mathscr{P}_n,
\ea
where $\delta \xi$ is the system precision (i.e. the grid-size) in measuring the distance. The $\text{\textdelta}(x)$ is defined such that $\lcb \text{\textdelta}(x) =0 \ | \ \forall x \neq 0 \rcb$ and $\text{\textdelta}(0) =1$.
In our reward signal definition, the first term reflects how well the agent is following (or catching) an assigned target point, while the second term penalizes any unnecessary speed-ups or -downs.
This penalty is assigned because the dipole strength of the swimmers is directly proportional to their swimming speed ($\mathscr{D} \sim v_sl_s^2$), and thus any change in the value of swimming speeds can disturb the balance of dipole strengths. However, the relative value of these speed modifications is kept negligibly small (here e.g. $\delta v/v^0 = 0.05$) while the agents are actively cloaking an intruder (i.e. robustly following it in a predefined arrangement).
Contribution of the third term in the reward signal ($\mathscr{C}_n$) is two-folded: (i) it encourages the agent to accurately get into the assigned target point once fairly close (here e.g. $\mathscr{C}_n = 100$ when $|\xvec^t-\xvec| \leq \delta \xi$), and (ii) it strictly penalizes wandering of the agent far off the target (here e.g. $\mathscr{C}_n = -100$ when $|\xvec^t-\xvec| \geq \Delta \xi$). 
Finally the last term, $\mathscr{P}_n = - \cos ^{-1} (\bm{e} \cdot \bm{e}^i)$, ensures that the agents learn how to smartly collaborate in a way that their induced disturbing flows do not disturb the cloaked object (i.e. the intruder) from its original path ($\bm{e}^i$) -- hence, realizing \textit{non-invasive} active cloaking. 

It is to be noted that realizing an active cloak around the assigned intruder, as the ultimate goal of this learning algorithm for each agent, is mathematically encoded as achieving maximal long-term \textit{accumulated} rewards. Therefore, negative nature of the reward signal $r_n$ \eqref{eq4.4} further ensures that our cloaking agents learn the shortest paths toward positioning themselves in the desired cloaking arrangement -- as any unnecessary action will result in an extra accumulation of negative rewards.

\begin{figure}
	\centering 
	\includegraphics[width=0.92\textwidth]{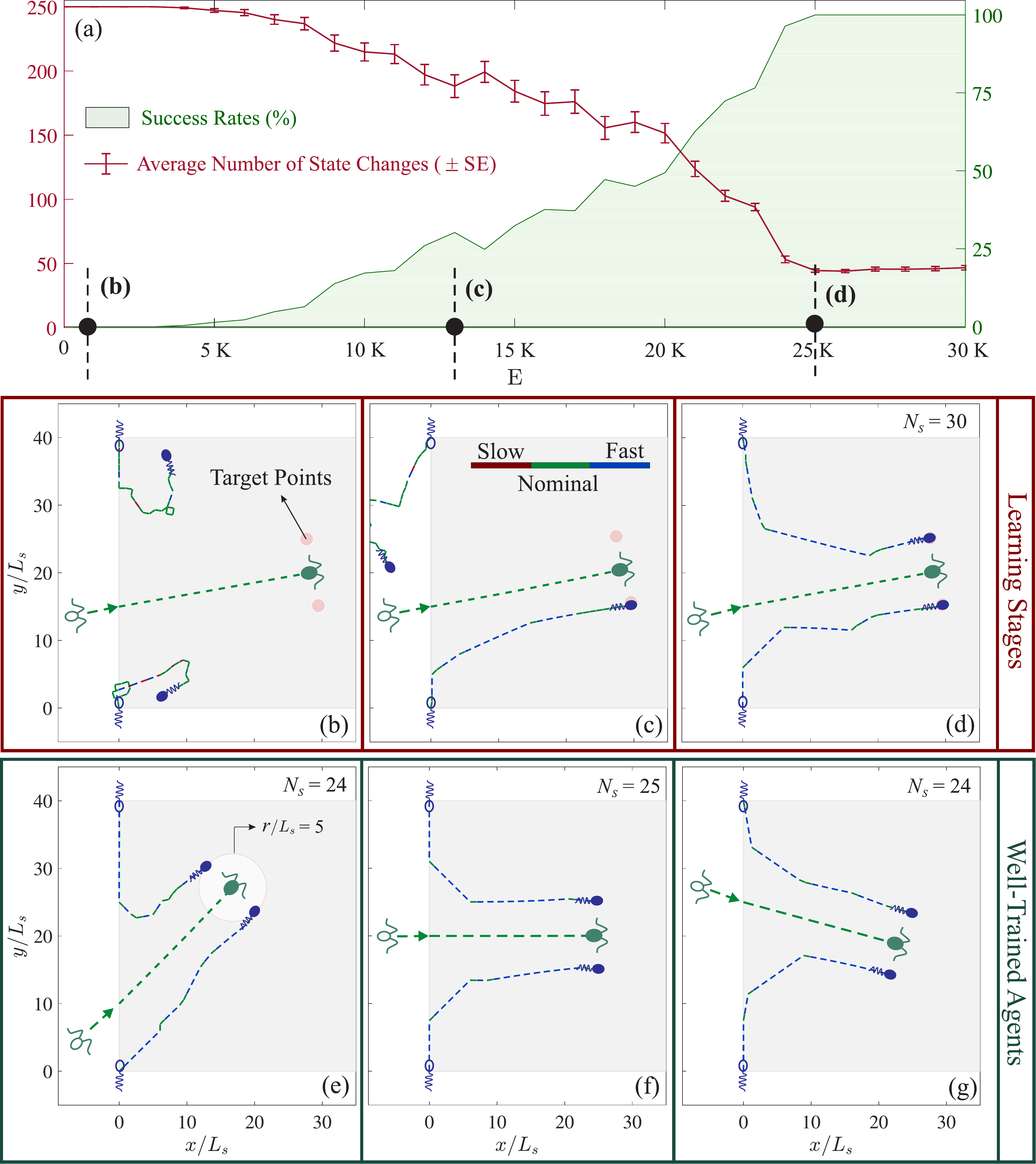}
	\caption 
	{\textbf{(a)} The learning process is assessed in terms of the agent's success rate, and number of the required swimming actions. After every $1000$ training episodes, a set of $100$ random active-cloaking tests have been performed using a purely deterministic policy based on the most updated $Q$-matrix at the moment. Each of the testing episodes starts with an intruder randomly entering a guarded region (shaded in gray; see panels b-g) from a random position and swimming toward a random direction, while the cloaking agents are initially positioned on the sides of entrance (see e.g. panels b-g). The test results are presented throughout the learning phase in terms of: (i) the success rate (shown in green) of the agents in catching the intruders and forming the desired active cloak, as well as (ii) the number of total state-changes (average $\pm$ Standard Error) in each set of testing episode (shown in red). \textbf{(b)}-\textbf{(d)}  Benchmarks of different learning stages. The panels represent a sample test on realizing an active cloak around a random intruder, performed in three different learning stages, i.e. non-adaptive (b), intermediate-adaptive (c), and well-adaptive (d) stages. The episode number ($E$) corresponding to each of the panels (b-d) is marked in panel (a), and agents use the most updated behavioral policy at each stage. The intruder (here a puller shown in green) and the cloaking agents (pushers shown in blue) are also represented both at the initial and final positions with stripe and solid schematics, respectively. Trajectories of the swimmers are shown in each panel by dashed lines, and color-coded based on the swimming speed at the moment -- see the legend in panel (c). The assigned target points (to realize the desired symmetrical active cloak) for each of the agents is also marked in each panel (pink markers). \textbf{(e)}-\textbf{(g)} Sample tests performed using well-trained cloaking agents, equipped with well-adaptive behavioral policy, which is obtained after convergence of the success rate to $100\%$ (see panel a). The number of required swimming actions (or equivalently the number of state-changes, $N_s$) is denoted on each panel. At this stage, the agents are capable of identifying the shortest non-invasive paths in forming desired active cloaks around any randomly moving intruder. The target arrangement here is to symmetrically position on the sides of intruder, with a separation distance $r_0/L_s =5$ (see e.g. panel e). The learning hyper-parameters are set to $\alpha=0.3$, $\gamma=0.95$, and $\epsilon=0.01$.}
	\label{figM3}
\end{figure}

Here, the cloaking agents are trained in pairs, and we employ a shared policy approach among them to accelerate the learning process.  
Training is conducted through consecutive learning episodes, denoted by $E = 1,2, \dots, n_E$, where $n_E$ is the total number of episodes in the learning phase. 
Each episode starts with an intruder randomly entering a guarded (sensitive) region -- i.e. from a random position and toward a random direction. The cloaking agents are also initially located on either side of the entrance (see Fig. \ref{figM3}). 
The first episode ($E=1$) is initialized by zero entries for all elements of the $Q$-matrix (i.e. an optimistic initialization) to further encourage exploration of new solutions and avoid trapping in local optima -- recall the negative nature of the reward signal. 
For the subsequent episodes ($E = i+1$, $i\geq1$), the action-value function, $Q$, will be initialized by that obtained at the end of previous episode (i.e. $E=i$).
The end of each training episode is declared when: either (i) the agents have formed a desired cloak around the intruder -- i.e. when they both have reached to the assigned target points in a predefined arrangement (see e.g. Fig. \ref{figM3}-d); or (ii) a certain number of state changes has passed without any success (e.g. $N_{s, \max}=250$ in Fig. \ref{figM3}a).
The process is in turn repeated until the action-value function coverages to an optimal value (i.e. $Q_n \rightarrow Q^*$ as $n_E \rightarrow \infty$) -- or practically, when the policy ($\pi_n$) converges to the optimal behavioral policy ($\pi_n \rightarrow \pi^*$).

To assess the learning process at different stages, we run a set of $100$ random active-cloaking tests throughout the training phase -- here, specifically after every $1000$ learning episodes. In testing the agents' intelligence, we deploy a purely deterministic (greedy) policy (i.e. $\epsilon=0$) based on the most-updated acquired $Q$-matrix. During each of these testing episodes, the agents are subject to cloak a random intruder using their so-far gained experience (encoded in the given $Q$-matrix). 
For each set of the ($100$) performed tests, we then compute the average number of total state-changes ($N_s$), along with success rate of the agents in catching the intruders and forming desired active cloaks. Both measures are monitored throughout the learning phase as demonstrated in Fig. \ref{figM3}(a).
The learning is evident in the gradual increase (decrease) of the agents' success rate (required number of swimming actions) in collaborative formation of the desired active cloak around random intruders (Fig. \ref{figM3}). 
As a benchmark, we also demonstrate, in panels (b)-(d) of Fig. \ref{figM3}, a specific sample test performed in three different learning stages -- corresponding episode numbers ($E$) are marked in panel (a).
While the agents are too naive at early stages (e.g. Fig. \ref{figM3}-b) and fail to catch and cloak the subject, they eventually learn how to actively cloak the moving intruder through an optimal set of non-invasive actions (e.g. Fig. \ref{figM3}-d).
Note that both of the monitored measures (i.e. the success rate and the number of swimming actions required to form a desired cloak) eventually converge to their optimal values as the number of learning episodes increases sufficiently (here, after $E \sim 25,000$).
The converged value of $100\%$ success rate for the cloaking agents, clearly shows their ability to smartly form a non-invasive active cloak around any randomly moving intruder (see e.g. the samples presented in Fig. \ref{figM3}e-g). In doing so, they find shortest non-invasive paths using only their own adaptive decision-making intelligence -- hence, the name \textit{smart} micro-swimmers.

\begin{figure}
	\centering 
	\includegraphics[width=0.85\textwidth]{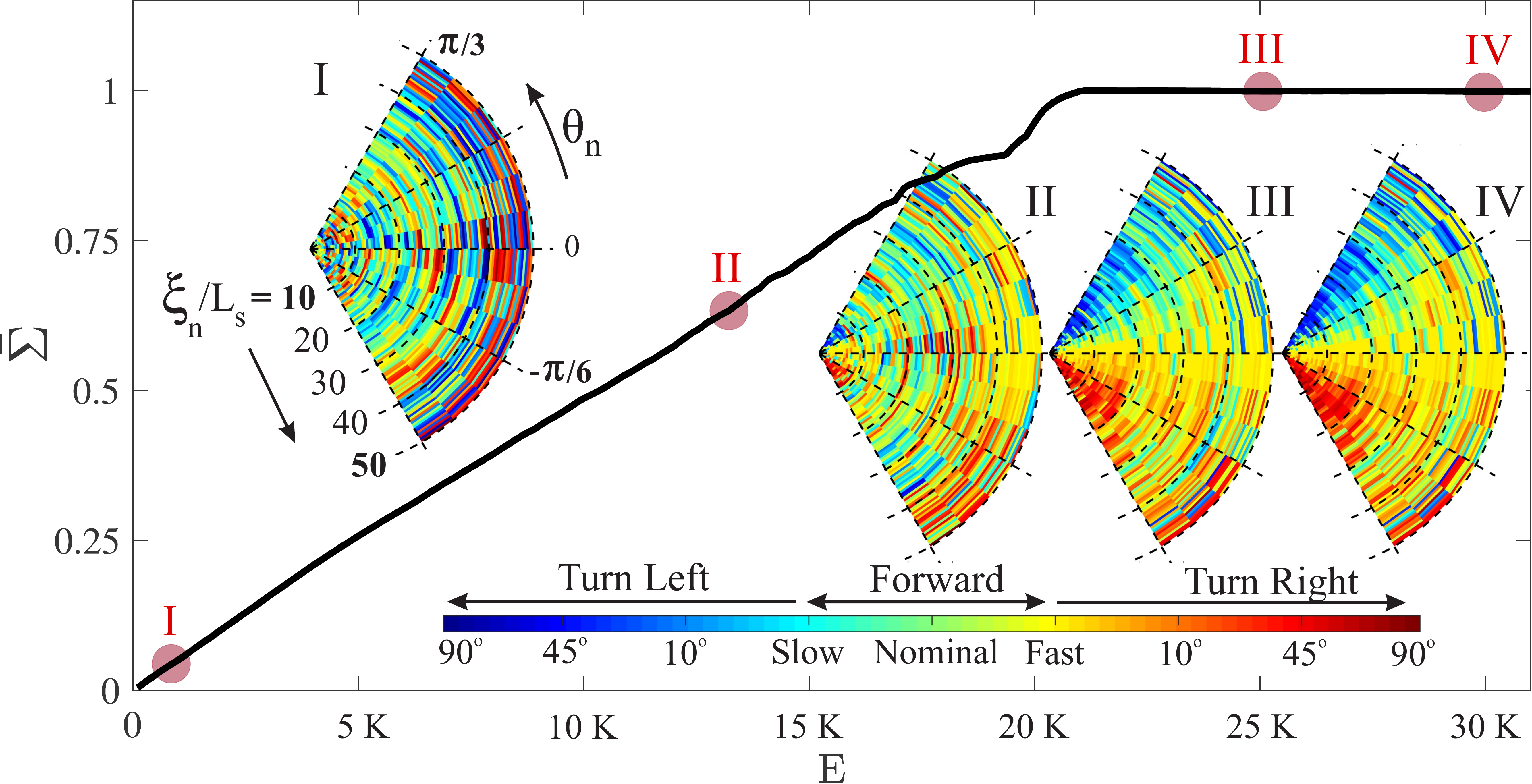}
	\caption {Evolution of the normalized element-wise sum of the $Q$-matrix ($\bar{\Sigma}$) through successive training episodes ($E$) during the learning phase. The agents' behavioral policy is also visualized (see insets I-IV) by color-coding over a selected sector of the state-space. Its evolution is monitored at four different learning stages (marked over the curve by I-IV), and eventually converges to the optimal behavioral policy. Here, each swimming action is presented by a color-code (see the color-bar legend), and every point in the state-space is shaded by a color representing the agents' current understanding of the best action at that specific state. The training episodes are the same as those presented in Fig. \ref{figM3}, for which the learning hyper-parameters are set to $\alpha=0.3$, $\gamma=0.95$, and $\epsilon=0.01$.}
	\label{figM4}
\end{figure}

Convergence of the cloaking agents' adaptive decision-making intelligence (encoded in the $Q$-matrix) is better demonstrated in Fig. \ref{figM4}, where we show evolution of the element-wise sum of the $Q$-matrix ($\Sigma$), normalized by its converged value. 
Additionally, the experientially learned behavioral policy is visualized by color-coding over a sector of the state-space, in four different learning stages (marked in Fig. \ref{figM4} by I-IV). Here, convergence of the action-value function (and thus the behavioral policy) is achieved when the number of successive training episodes ($E$) exceeds $\sim 25,000$ (Figs. \ref{figM3}-\ref{figM4}).

Once the learning process is converged, the cloaking agents are readily equipped with an \textit{optimal} behavioral policy. 
To be more precise, the optimal behavioral policy refers to the policy ($\pi^*: s \rightarrow a$) that in any given state ($s$), determines the action ($a$) which maximizes the expected sum of future rewards (encoded in $Q^*$). 
Here, the reward signal ($r_{n+1}$) received (during the learning process) by the agents after taking an action ($a_n$), has two salient features -- c.f. Eq. \eqref{eq4.4}: (i) it reflects a penalty for disturbing the intruder from its original path, and (ii) it is designed to always have a negative value, except at the target state. 
Therefore, maximizing the discounted sum of future rewards will ensure, not only to successfully realize the desired non-invasive cloak, but also to do so with minimum number of actions (Fig. \ref{figM3}a). Note that spending more time (by taking excessive actions) will simply result in accumulating more negative rewards, which is discouraged by the optimal policy.  
This enables the agents to find \textit{shortest} non-invasive paths toward positioning themselves in desired cloaking arrangements around arbitrarily moving intruders (see e.g. the samples presented in Fig. \ref{figM3}e-g).

\begin{figure}
	\centering 
	\includegraphics[width=0.85\textwidth]{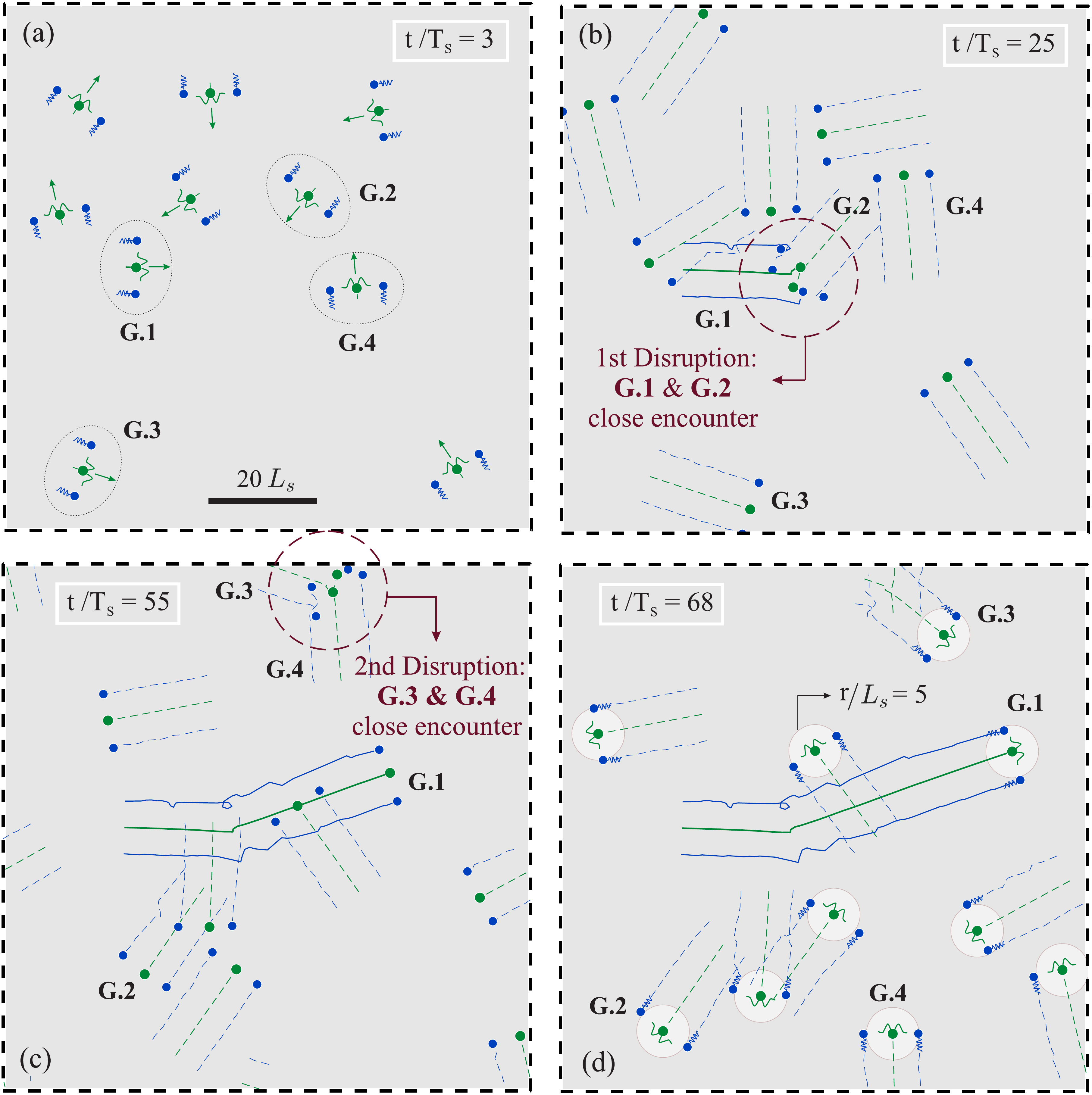}
	\caption {Time evolution of a crowded suspension of intruders (shown in green), each of which actively cloaked by a pair of smart micro-swimmers (shown in blue). Here the intruders are freely moving toward random directions in space -- specified by green arrows in (a). Periodic boundary conditions are imposed on the presented panels -- i.e. each panel represents just a window of an infinite domain at that specific moment. As demonstrated through snapshots (a)-(d) of the system's time evolution (see also Movie S1), using their adaptive decision-making intelligence, our agents are able to robustly maintain the cloak formation in the presence of complex hydrodynamic interactions. They are also able to immediately restore the desired formation after any sever close encounters (b-c) which cause major disruptions to the cloak. As a benchmark, the evolution dynamics of four sample groups (G.1-4) are tracked in the panels. Specifically, the close encounter between G.1 \& G.2 as well as G.3 \& G.4 are marked by red dashed circles in panels (b) and (c), respectively. Also, trajectories of the swimmers in G.1 are shown by solid lines throughout the time, while other groups' trajectories are only demonstrated (by dashed lines) within the last $25T_s$ of their motion. The time corresponding to each snapshot is noted in each panel, and swimmer schematics are added to panels (a) and (d) for readability. Here, the goal was to realize an active version of the symmetric cloak presented in Fig. \ref{figM1}. Thus, for the reference, rings of radii $r/L_s=5$ (that is equal to the predefined desired separation distances) are depicted around intruders in the final snapshot (d).}
	\label{figM5}
\end{figure}

However, there is still a caveat here: the presence of long-range flow-mediated interactions, if not responded adaptively by taking proper swimming actions, will soon distort the arrangement and reveal the cloaked object. 
Our swimming agents, however, have been trained not only to `\textit{catch}' and form desired cloaks around random intruders, but also to `\textit{follow}' them robustly, while maintaining the implemented cloak throughout their motion. 
As a benchmark, let us monitor the time evolution of a sample crowded suspension of intruders (Fig. \ref{figM5}), where the deployed pairs of our smart micro-swimmers have already formed the desired cloak around each of the subjects -- see the examples presented in Fig. \ref{figM3}.  
Periodic boundary conditions are imposed on the presented panels, as if they represent just a window of an infinite domain of suspension. 
Such a complex random system is mainly characterized by each of the swimmers moving toward a random direction in space (Fig. \ref{figM5}a), with frequent close encounters happening between them (e.g. Fig. \ref{figM5}b-c). 
It worths noting that even an isolated intruder, swimming (alone) on a straight line in an unbounded domain, can distort the arrangement of an implemented cloak, and cause naive cloaking agents to diverge or collapse. The situation here is further exacerbated by the presence of other nearby swimmers, given their disturbing flows and long-range flow-mediated interactions -- let alone close encounters happening in such a crowded system, and the consequent disruptions in cloak arrangements. 
However, as it is evident throughout the system evolution (see also Movie S1), the proposed \textit{smart} active cloaking strategy is robust. 
In particular, by taking optimal adaptive swimming actions, the cloaking agents not only successfully retain their arrangement in the presence of hydrodynamic interactions, but also immediately re-form the desired cloaking arrangements after any sudden disruption (Fig. \ref{figM5}, Movie S1).

\begin{figure}
	\centering 
	\includegraphics[width=0.93\textwidth]{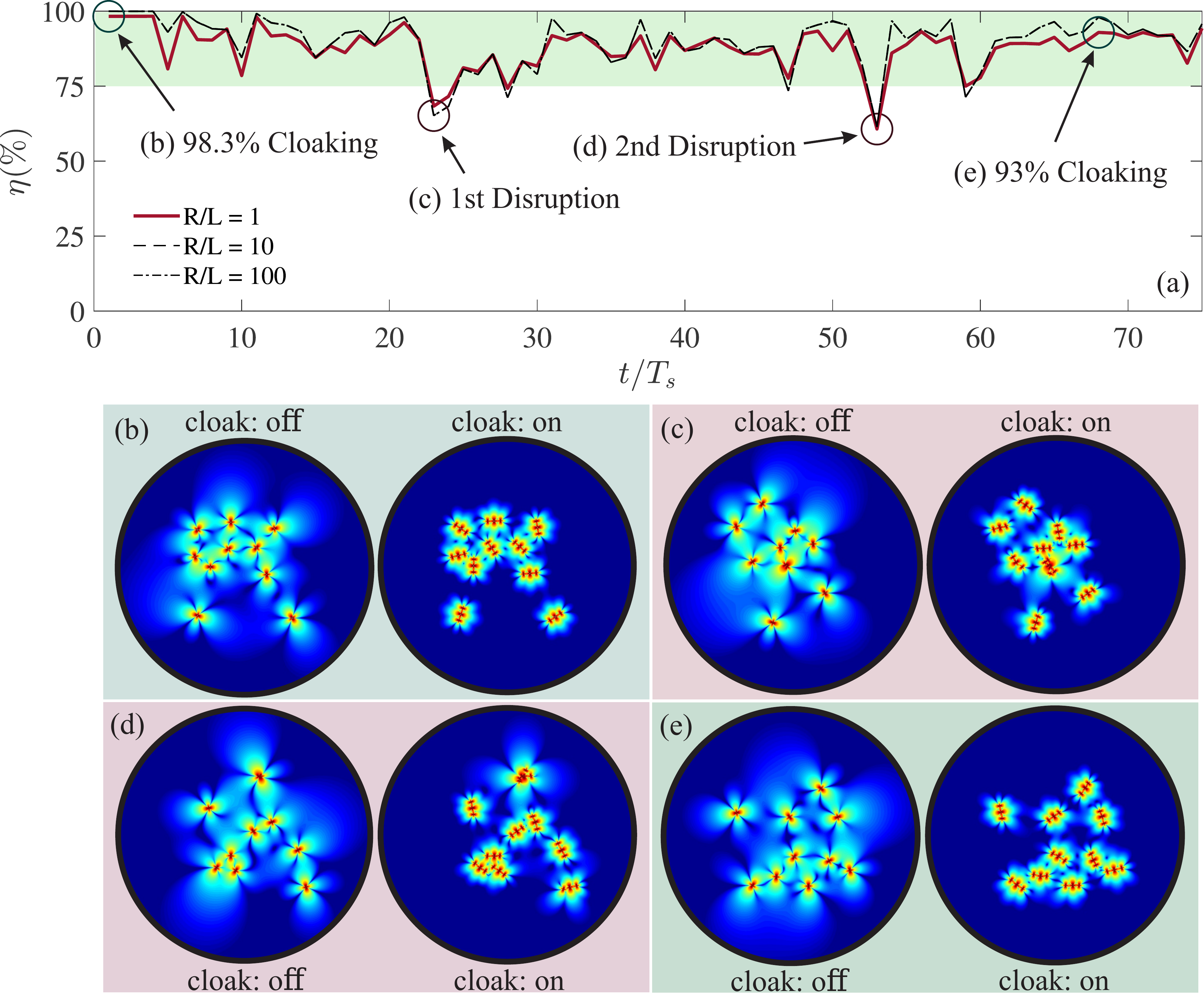}
	\caption {\textbf{(a)} Time evolution of the overall cloaking performance ($\eta\%$) corresponding to the flock of smart micro-swimmers deployed within the crowded suspension presented in Fig. \ref{figM5} to cloak random intruders. Here $\eta$ ($\%$) is monitored over the entire time evolution at ranges $R/L=$ 1 (red solid line), $R/L=$ 10 (black dashed line), and $R/L=$ 100 (black dash-dotted line), where $L$ is the length of each presented window in Fig. \ref{figM5}. \textbf{(b-e)} Snapshots of the induced fluid disturbances (see also Movie S1) are visualized (by color shading) over the entire system (cloak: on), and also compared to those induced when no cloaking agents are deployed (cloak: off). The moment corresponding to each of the snapshots in (b-e) is marked on the curve in (a). Specifically, we demonstrate the snapshots at: (i) an early stage, where the desired cloaks ($\eta \approx 98\%$) have been just formed (Figs. \ref{figM5}a, \ref{figM6}b); (ii) the moments of close encounters happening between G.1 \& G.2 (1st Disruption, see Figs. \ref{figM5}b, \ref{figM6}c) as well as G.3 \& G.4 (2nd Disruption, see Figs. \ref{figM5}c, \ref{figM6}d), respectively, where the implemented cloaks undergo sever disruptions; and (iii) a final recovered stage (Figs. \ref{figM5}d, \ref{figM6}e) where the agents are once again perfectly in the desired cloaking arrangements ($\eta \approx 93\%$).}
	\label{figM6}
\end{figure}

In practice, it is often of interest to conceal a single micro-swimmer (e.g. a biological micro-robot) within a crowded suspension (e.g. inside the human body) using only a pair of cloaking agents. 
However, for the sake of generality, here we have used a \textit{flock} of cloaking agents to cloak all the (randomly moving) swimmers in the presented crowded suspension (see Fig. \ref{figM5}). 
To assess performance of the deployed flock in concealing the assigned swimming objects, we then monitor the overall cloaking efficiency ($\eta\%$) throughout the system's time evolution (Fig. \ref{figM6}a).
The magnitude of disturbing flows (i.e. flow signatures) induced by the system is also monitored throughout the represented time evolution (Movie S1), and is compared to the same system when no cloaking agents are deployed (Fig. \ref{figM6} b-e).

The implemented cloaks are dynamically adjustable, as the cloaking agents behave adaptively in response to fluid-mediated interactions. This enables them to robustly keep their specified arrangement, and actively conceal any moving subject throughout its motion (see e.g. Fig. \ref{figM5}).
Not to mention that the agents first need to smartly collaborate in pairs and find optimal paths to separately catch each of the assigned subjects (see the samples in Fig. \ref{figM3}) forming an active set of perfect cloaks (with $\eta > 98 \%$) around these randomly moving intruders (Figs. \ref{figM5}a, \ref{figM6}b). 
Once realized, throughout the system evolution, the cloaking efficiency remains well above 75\% even in the presence of non-linearly varying complex flow-mediated interactions (Fig. \ref{figM6}a). The two relatively notable sudden drops in the cloaking performance (i.e. to $\eta \sim$ 70\%) occur right after the reported close encounters, which cause major disruptions to the cloaks' arrangements (Fig. \ref{figM5} b-c, Fig. \ref{figM6} c-d). However, the implemented active cloaks are robust, thus are immediately re-formed and the cloaking efficiency is quickly recovered. For instance, note the final stage (Figs. \ref{figM5}d, \ref{figM6}e) with a remarkable $93\%$ cloaking efficiency, that is recovered after going through many interactions and encounters.

It is noteworthy that the deployed flock of cloaking agents have never been trained for, neither exposed to, this (or any other) specific scenario with \textit{multiple} intruders. 
This means that once the agents undergo the described \textit{catch} and \textit{follow} training processes (see Fig. \ref{figM3}), they can be readily used in pairs to actively cloak any number of arbitrary intruders within unexposed crowded suspensions.

%%%%%%%%%%%%%%%%%%%%%%%%%%%%%%%%%%%%%%%%%%%%%%%%%%%%%%%%%%%%%%%%%%%%%%%%%%%%%%%%%%%%%%%%%%%%%%%%%%%%%%%%%%%%
\section{Concluding Remarks}

In this article, we presented a rigorous approach to actively cloak swimming objects in the Stokes regime using micro-swimmers equipped with adaptive decision-making intelligence.
Through a reinforcement learning algorithm, our cloaking agents experientially learn optimal adaptive behavioral policy in the presence of non-linear flow-mediated interactions.
This artificial intelligence enables them to dynamically adjust their swimming actions, so as to optimally form, and robustly retain, desired cloaking arrangements around any arbitrarily moving object. 
In doing so, our \textit{smart} agents optimally cooperate, such that their overall generated flows not only cancel out the cloaked object's induced fluid disturbances, but also do not disturb it from its original path. 
Therefore, the presented active cloaking approach is also \textit{non-invasive} to the subject's motion. 
We then generalized our methodology, and demonstrated that a cooperative flock of well-adapted cloaking agents can be readily used in a crowded environment to actively cloak any number of arbitrary intruders.

Our study provides a clear road-map toward realizing hydrodynamic invisibility cloaks for externally or internally controlled artificial swimming micro-robots \cite[e.g.][]{li2017micro}.
This will be significant in non-invasive intrusion of swimming micro-robots with a broad range of biomedical applications \cite[see e.g.][]{nelson2010microrobots}. 
Moreover, our findings demonstrate the great potential of reinforcement learning in paving the path toward engineering of smart micro-swimmers capable of accomplishing a new class of group-objectives. We, therefore, hope that this article will spur further research on this field at the intersection of fluid mechanics and artificial intelligence.

%%%%%%%%%%%%%%%%%%%%%%%%%%%%%%%%%%%%%%%%%%%%%%%%%%%%%%%%%%%%%%%%%%%%%%%%%%%%%%%%%%%%%%%%%%%%%%%%%%%%%%%%%%%%
\section*{Acknowledgments}

\no 
This work is supported by the National Science Foundation grant CMMI-1562871.

\section*{Declaration of Interests}

\no 
The authors report no conflict of interest.

%\renewcommand\thefigure{A\arabic{figure}}    
%\setcounter{figure}{0} 

%\newpage
%%%%%%%%%%%%%%%%%%%%%%%%%%%%%%%%%%%%%%%%%%%%%%%%%%%%%%%%%%%%%%%%%%%%%%%%%%%%%%%%%%%%%%%%%%%%%%%%%%%%%%%%%%%%

\section*{Appendix}

\subsection*{A.1. On the Generality of a Learned Policy}\hfill
\vspace{3mm}

As discussed in section \ref{sec.Learning}, the reinforcement learning process will equip our cloaking agents with an adaptive decision-making intelligence. 
A flock of these smart micro-swimmers can then be readily used to actively cloak any number of arbitrarily moving intruders, within any unexposed crowded suspension (see e.g. Figs. \ref{figM5}-\ref{figM6}).
However, it still remains unclear to what extent a learned policy can be effective once used: (i) by agents with different swimming speeds, and/or (ii) for cloaking intruders of various swimming speeds.

\begin{figure}
	\centering 
	\includegraphics[width=0.8\textwidth]{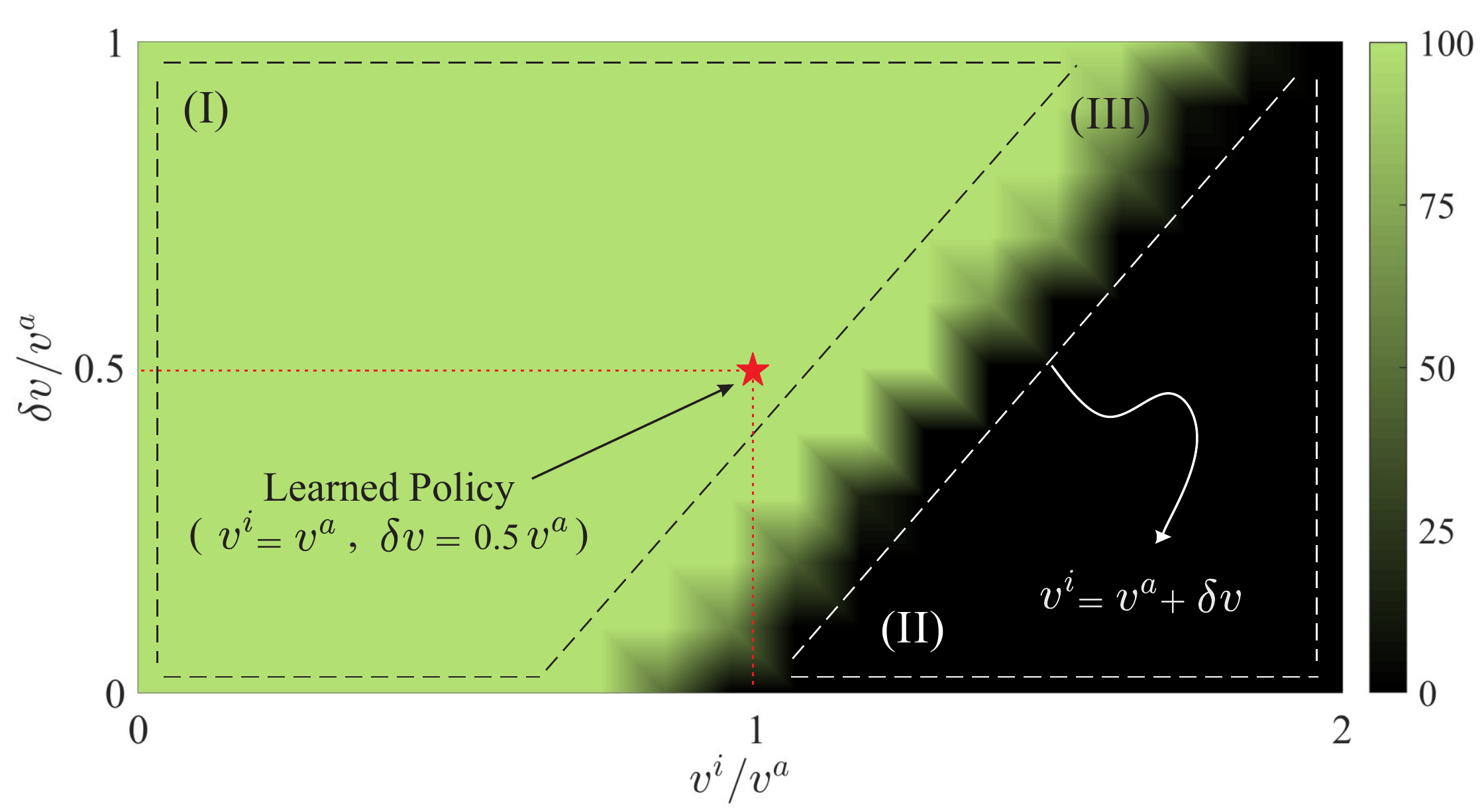}
	\caption {Applicability of a learned policy (measured in terms of success rate) in cloaking intruders with various swimming speeds ($v^i$), once adopted by agents with different speed characteristics (i.e. $v^a$ and $\delta v$). The employed policy ($\pi^*$) is specifically the one learned by agents with $\delta v/v^a = 0.5$, through episodes of catching intruders swimming in random directions with the speed of $v^i/v^a = 1$ (see Fig. \ref{figM3}). The reference point is marked (with a red asterisk) on the parameter-space ($v^i/v^a$, $\delta v/v^a$). To assess the generality of this policy, we run a set of $25$ random active-cloaking tests (similar to those demonstrated in Fig. \ref{figM3}) for every point in the presented parameter-space (i.e. corresponding to each specific choice of $v^i/v^a$ and $\delta v/v^a$). Color shading represents success rate of the deployed agents (equipped with the same $\pi^*$) in forming desired cloaks (i.e. the one with $\beta=\pi/2$ and $\psi = \pi$ in Fig. \ref{figM1}) around the corresponding randomly moving intruders. Three distinct regions can be identified in the parameter-space: (I) the region of ideal applicability within which the learned policy is fully effective with a success rate of 100\%, (II) the region associated with $v^i > v^a + \delta v$, for which catching the intruder is not physically possible, and (III) a transition region.}
	\label{figA1}
\end{figure}

Let us consider the optimal policy ($\pi^*$) obtained specifically for the agents with $\delta v/v^a = 0.5$, through episodes of catching intruders swimming in random directions with the speed of $v^i/v^a = 1$ (see e.g. Figs. \ref{figM3}-\ref{figM4}). 
The generality of this policy is assessed in Fig. \ref{figA1} over the parameter-space ($v^i/v^a$, $\delta v/v^a$).
In particular, for every point in the presented parameter-space (i.e. corresponding to each specific choice of $v^i/v^a$ and $\delta v/v^a$), we run a set of $25$ random active-cloaking tests (similar to those demonstrated in Fig. \ref{figM3}).
The results of our numerical experiments (presented in Fig. \ref{figA1}) reveal a wide range of ideal applicability (region I), within which the learned policy is fully effective (with a success rate of 100\%) in cloaking random intruders.
This almost covers the entire parameter-space, except the cases with $v^i > v^a + \delta v$ (region II), for which catching the intruder is not physically feasible, followed by a transition zone (region III).

%

%%%%%%%%%%%%%%%%%%%

\begin{figure}
	\centering 
	\includegraphics[width=0.95\textwidth]{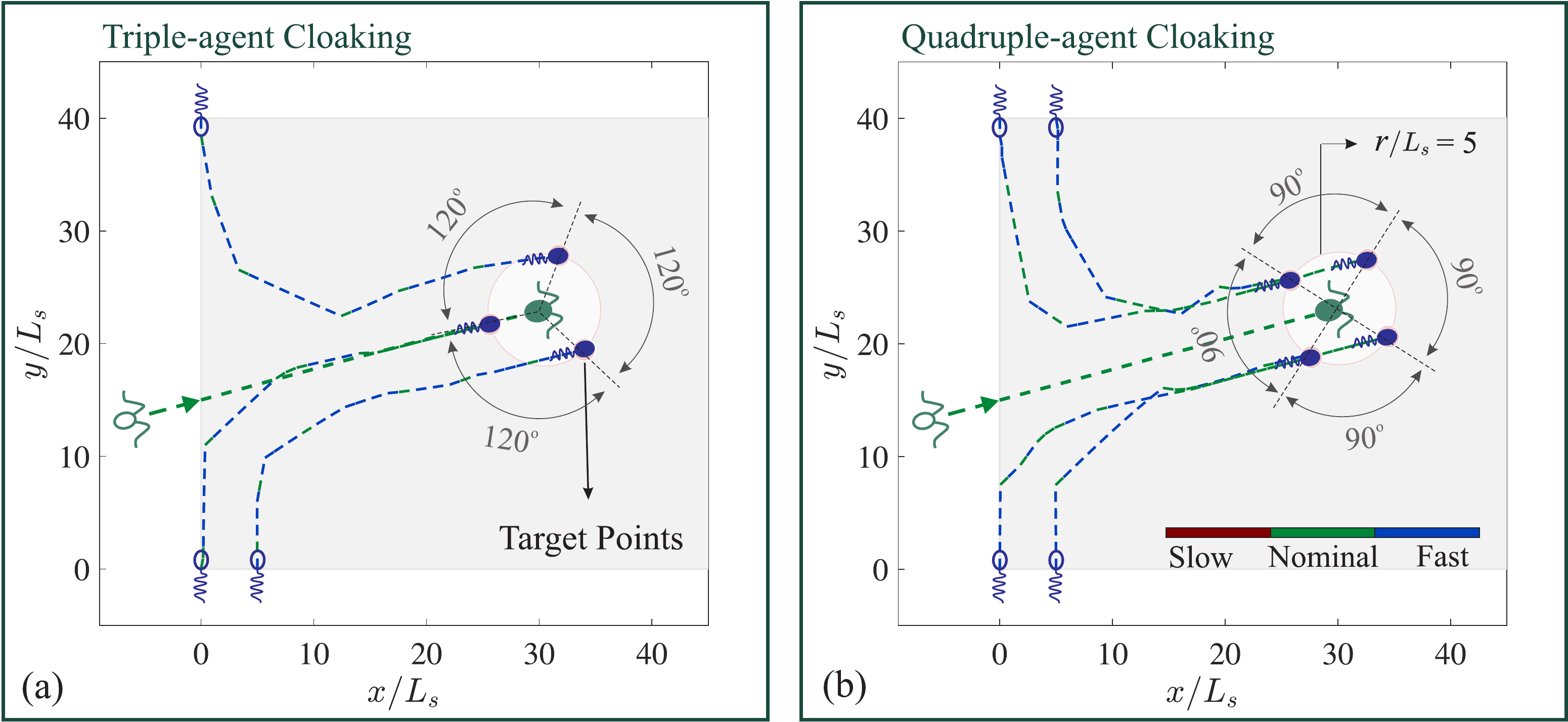}
	\caption {Applicability of a learned policy in realizing multi-agent active cloaks for random intruders. Here, the employed policy ($\pi^*$) is the one obtained through consecutive learning episodes of \textit{double-agent} cloaking, as demonstrated in Fig. \ref{figM3}. Once the agents are equipped with this behavioral policy, they can be readily deployed to form any other forms of cloaking arrangements around random intruders. Here, as a benchmark, we demonstrate samples of triple- and quadruple-agent cloaking strategies in panels (a) and (b), respectively. Similar to those presented in Fig. \ref{figM3} (b-g), each testing episode starts with an intruder entering a guarded region (shaded in gray) from (toward) a random position (direction), while the cloaking agents are initially positioned on the sides of entrance. The intruder (here a puller, shown in green) and the cloaking agents (pushers shown in blue) are also represented both at the initial and final positions with stripe and solid schematics, respectively. Trajectories of the swimmers are shown in each panel by dashed lines, and color-coded based on the swimming speed at the moment -- see the legend in panel (b). The assigned target points for each of the agents (to realize the desired symmetrical active cloaks) are also marked in each panel (with pink markers). Note that the agents (equipped with optimal behavioral policy) are capable of identifying the shortest non-invasive paths toward forming any desired active cloak around any randomly moving intruder. The target arrangements here are symmetric configurations, achieved by positioning three (a) or four (b) agents around the subject, with a separation distance of $r_0/L_s =5$ (denoted by a pink circle around the intruder).}
	\label{figA2}
\end{figure}

Our numerical experiments also reveal that the agents equipped with a learned behavioral policy, can be readily deployed to realize any other form of cloaking arrangements around a random intruder.
Here, as a benchmark, we demonstrate how our smart agents can be readily used in \textit{triple-} or \textit{quadruple-agent} cloaking strategies (see Fig. \ref{figA2}).
We once again note that when a pair or multiple micro-swimmers perform active cloaking, the state space is defined for \textit{each} agent independently. To be more precise, the state space for each agent includes its distance and relative orientation with respect to the intruder, while all the agents use a shared action policy.
It is to be highlighted that the policy ($\pi^*$) adopted by agents in the presented triple- and quadruple-agent cloaking test cases (Fig. \ref{figA2}), is the one obtained through consecutive learning episodes of \textit{double-agent} cloaking (presented in Fig. \ref{figM3}). 
Therefore, the deployed agents have no prior experience in forming such three- or four-agent cloaks.

%

%%%%%%%%%%%%%%%%%%%%%%%%%%%%%%%%%%%%%%%%%%%%%%%%%%%%%%%%%%%%%%%%%%%%%%%%%%%%%%%%%%%%%%%%%%%%%%%%%%%%%%%%%%%%
\subsection*{A.2. On the Design of a Reward Signal}\hfill
\vspace{3mm}

In a reinforcement learning process, the actions taken by an agent are rewarded with a feedback signal ($r_n$) that quantifies its immediate success in achieving a predefined objective.
This form of feedback on performance is the only information that agents receive throughout their interaction with the environment. 
For them, the learning objective is to maximize the sum of accumulated rewards, which should properly reflect the ultimate \textit{physical} objective.
This makes the design of a proper reward signal a challenge for any reinforcement learning process.

Here, the ultimate goal for the agents is to obtain the ability to identify optimal (i.e. shortest) non-invasive pathways toward positioning themselves into desired cloaking arrangements, around any arbitrarily moving intruder.
In section \ref{sec.Learning}, we devised a reward function \eqref{eq4.4} composed of four elements, each meticulously encoding a particular aspect of this objective into the feedback signal received by the agents. 
Here, we further explore robustness of the presented learning approach against excluding each of those elements from the proposed signal.
To this end, let us consider the following set of alternative reward functions:
\bsa 
&r_n^{(1)} = [1/(\xi_n+\delta \xi) - \xi_n] + [\text{\textdelta}(v_n-v^0)-1] + \mathscr{C}_n + \mathscr{P}_n \ , \label{eq.A1a}\\
&r_n^{(2)} = [1/(\xi_n+\delta \xi) - \xi_n] + [\text{\textdelta}(v_n-v^0)-1] + \mathscr{C}_n \ , \label{eq.A1b}\\ 
&r_n^{(3)} = [1/(\xi_n+\delta \xi) - \xi_n] + [\text{\textdelta}(v_n-v^0)-1] \ , \label{eq.A1c}\\
&r_n^{(4)} = [1/(\xi_n+\delta \xi) - \xi_n], \label{eq.A1d}\\
&r_n^{(5)} = -1 \ ,\label{eq.A1e}
\esa 
where $r_n^{(1)}$ is the original reward signal (used throughout this work), and $r_n^{(2)}$ to $r_n^{(5)}$ are the ones obtained through systematic elimination of the constitutive elements from $r_n^{(1)}$.
As discussed in section \ref{sec.Learning}, the first term of the original reward signal (i.e. $r_n^{(1)}$) reflects how well each agent is following (or catching) an assigned target point.
The second term is to penalize any unnecessary speed-ups or -downs by the agents, and thereby further encourages them to keep the dipole-strength in balance.
The third term in the signal has two main contributions: (i) it encourages each agent to get into the precise position of an assigned target point (once in its close proximity), and (ii) it strictly penalizes wandering of the agents far away from their target points. 
Finally, the last term ensures that the agents learn how to smartly collaborate in a way that their induced flow fields do not disturb the cloaked object (i.e. the intruder) from its original path -- hence, realizing \textit{non-invasive} active cloaking. 
We also note that the negative nature of the presented reward signals further ensures that our cloaking agents learn the \textit{shortest} paths toward positioning themselves in the desired cloaking arrangement -- as any unnecessary action will result in an extra accumulation of negative rewards.

\begin{figure}
	\centering 
	\includegraphics[width=0.75\textwidth]{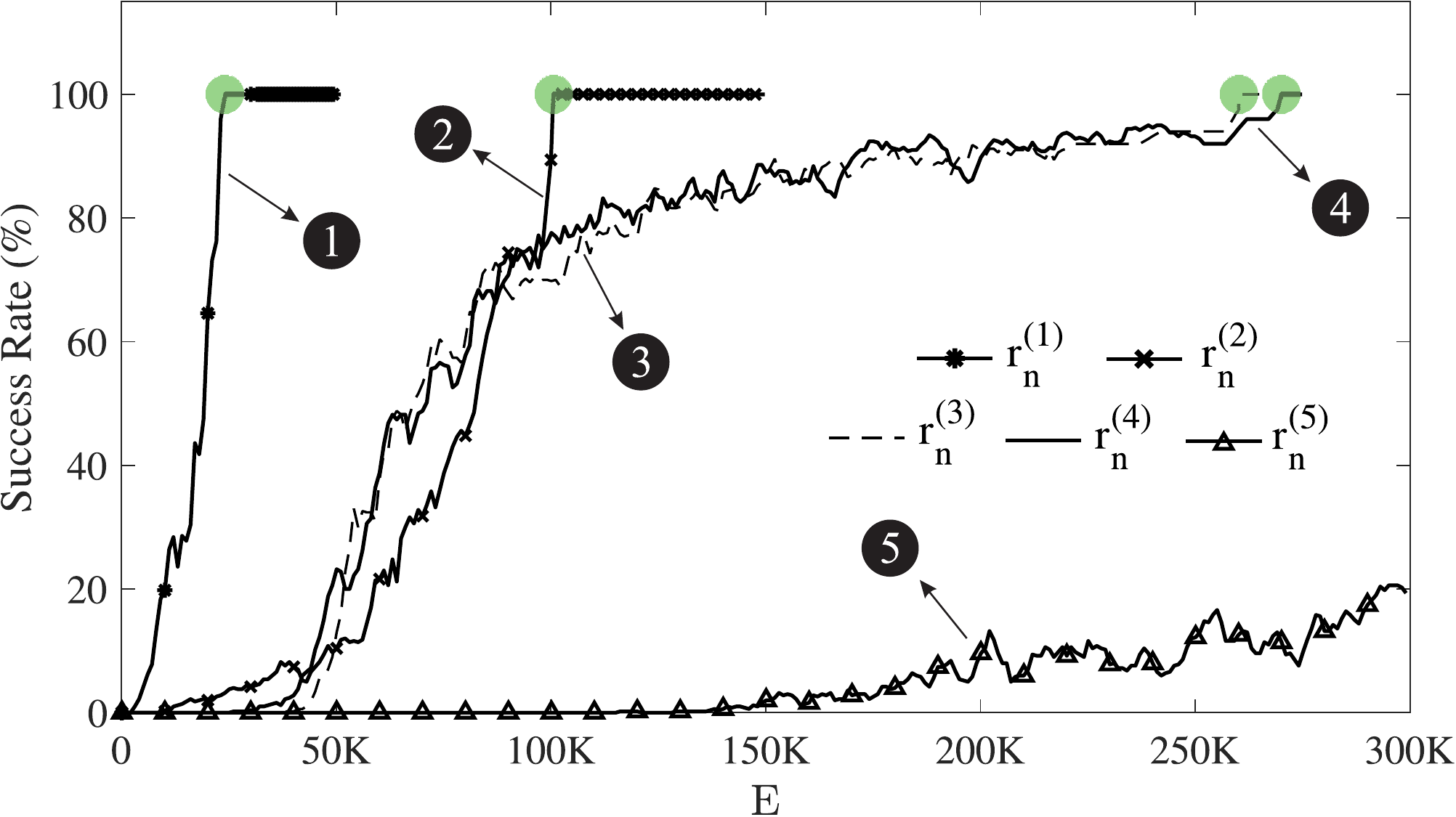}
	\caption {The learning process repeated for each alternative reward signal, $r_n^{(1)}$ to $r_n^{(5)}$, is assessed in terms of the agents' success rate. Similar to the original process demonstrated in Fig. \ref{figM3}, a set of $100$ random active-cloaking tests have been performed after every $1000$ training episodes, where the agents use a purely deterministic (greedy) policy based on the most updated $Q$-matrix at the moment. Results are presented for each case throughout the corresponding learning process, in terms of the success rate of the agents in catching the random intruders and forming desired active cloaks. The target arrangement is defined similar to the original process, that is the symmetric double-agent cloaking with a separation distance of $r_0/L_s =5$. Learning hyper-parameters are also kept the same -- i.e. $\alpha=0.3$, $\gamma=0.95$, and $\epsilon=0.01$. Here, the learning processes represented by curves (1)-(4) have converged (as marked in green) to the 100\% success rate after $\sim$ 25000, 100000, 260000, and 270000 training episodes, respectively. Note that curve (1) is the originally proposed learning process, also presented in Fig. \ref{figM3}(a).}
	\label{figA3}
\end{figure}

Here we repeat the learning process, using each of the proposed alternative feedback signals.
Similar to the original process (demonstrated in Fig. \ref{figM3}), alternative learning processes are then independently assessed (Fig. \ref{figA3}) via running random sets of $100$ active-cloaking tests after every $1000$ training episodes.
The results further suggest robustness of the presented methodology against excluding different elements from the devised reward signal ($r_n^{(1)}$).
However, there is a caveat here: each of the described components, are included to quantify the agents' success in achieving a particular aspect of their ultimate goal.
Failing to provide an explicit feedback on any of these aspects, makes the learning process substantially longer, and thus significantly increases the computational costs.
For instance, by eliminating the last term from the reward signal $r_n^{(1)}$, the agents will no longer receive an explicit (negative) feedback when disturbing the intruder. As a result, it takes a significantly larger number of trial-and-errors (i.e. training episodes) for them to learn that such a behavior is sub-optimal -- as it will increase the number of actions (and thus the time) required to catch an intruder.
Therefore, performance of the agents trained by the reward signal $r_n^{(2)}$ requires $\sim$ 100,000 learning episodes to converge to the 100\% success rate (Fig. \ref{figA3}), which is significantly higher than the $\sim$ 25,000 episodes required for those trained by $r_n^{(1)}$.
Similarly, the learning processes corresponding to the agents trained by further truncated signals $r_n^{(3)}$ and $r_n^{(4)}$, converge to the full performance rate after $\sim$ 260,000 and 270,000 episodes, respectively (Fig. \ref{figA3}).

We once again highlight that using an appropriately devised reward signal will significantly facilitate the learning process (see e.g. Fig. \ref{figA3}).
In fact, as the reward signal (received by the agents) becomes less informative about the ultimate objective, the required number of training episodes (i.e. the computational cost) for the agents to learn optimal behavioral policy grows substantially.
As an extreme case, let us consider $r_n^{(5)}$, where all components are eliminated from the originally proposed reward signal, keeping only the negative nature of the feedback.
This means that throughout the learning process agents will merely receive a constant reward of `$-1$' after any action they take, regardless of the outcome. 
Swimming agents trained by this naively defined feedback on performance, still can potentially learn how to realize desired active cloaks -- as it is the only way for them to stop accumulating more negative rewards.
However, the process is cumbersome, reaching to only $\sim$ 20\% success rate after 300,000 training episodes (Fig. \ref{figA3}), and requires a few million to achieve the full performance.

%%%%%%%%%%%%%%%%%%%%%%%%%%%%%%%%%%%%%%%%%%%%%%%%%%%%%%%%%%%%%%%%%%%%%%%%%%%%%%%%%%%
\subsection*{A.3. Alternative Control Strategies}\hfill
\vspace{3mm}

In section \ref{sec.Learning} of this article, we presented a systematic methodology to equip cloaking agents with an adaptive decision-making intelligence in response to flow-mediated interactions.
We then described how such an artificial intelligence encodes consequences of each possible action (at any given state) into the learned behavioral policy ($\pi^*$), based on which optimal actions are taken.
This makes the agents inherently aware of the shortest non-invasive paths (from anywhere in the state space) toward positioning themselves into a desired cloaking arrangement around an arbitrarily moving intruder.

\begin{figure}
	\centering 
	\includegraphics[width=0.95\textwidth]{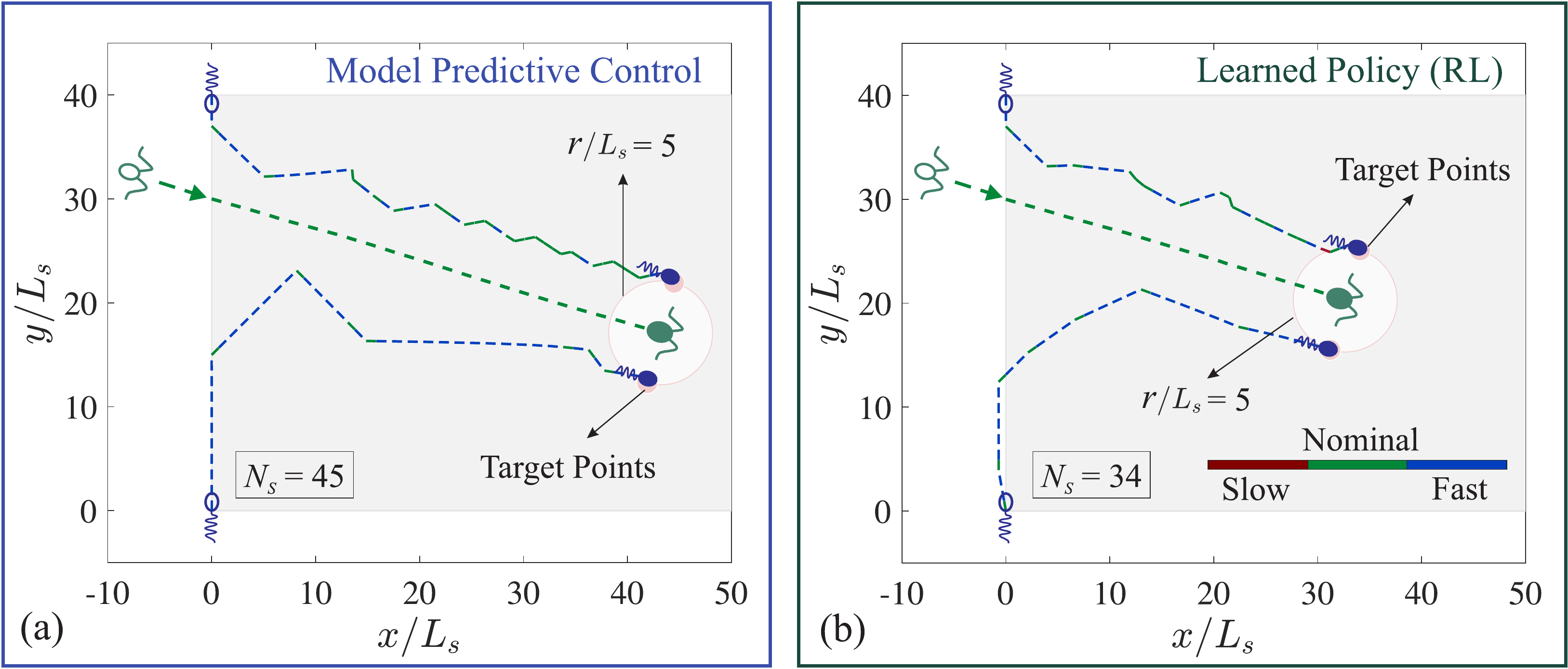}
	\caption {
		Performance of the implemented model predictive control (as an alternative agent-control strategy) is compared on a benchmark test case, against the performance of smart agents equipped with the optimal behavioral policy ($\pi^*$) obtained through the proposed reinforcement learning-based algorithm. Similar to the sample test cases presented in Fig. \ref{figM3} (b-g), the testing episode starts with an intruder entering a guarded region (shaded in gray) from (toward) a random position (direction), while the cloaking agents are initially positioned on the sides of entrance. Throughout the process, cloaking agents in panel (a) are actively controlled through a model predictive control scheme that targets the pre-identified optimal double-agent cloaking configuration. On the other hand, the agents deployed in panel (b) use their own adaptive decision-making intelligence encoded in the learned optimal policy ($\pi^*$), to realize the desired cloak. In both panels, the intruder (here, a puller shown in green) and the cloaking agents (pushers shown in blue) are represented both at their initial and final positions, with stripe and solid schematics, respectively. Trajectories of the swimmers are shown in each panel by dashed lines (color-coded based on the instantaneous swimming speed). The number of swimming actions (or equivalently, the number of state-changes $N_s$) taken in each case to form the desired cloak around the intruder is also reported within each panel.
	}
	\label{figA8}
\end{figure}

Alternatively, one may use an active optimal control strategy to guide the deployed swimming micro-robots (i.e. the cloaking agents) toward the desired arrangement.
Here, as a benchmark, we implement a model predictive control strategy \cite{camacho2013model} based on the same state-action space ($s_n, a_n$) used in the proposed reinforcement learning algorithm (see section \ref{sec.Learning} for details). 
In this alternative approach, the swimming agents are actively controlled to target the optimal double-agent cloaking configuration (already identified in section \ref{sec.Cloaking}). 
In doing so, the controlled agents are provided (throughout the process) with a feedback signal informing them about their current state, and then use a predictive model (that incorporates agent/intruder hydrodynamic interactions) to take the best action at any given state.
We compare the performance of this alternative agent-control strategy on a benchmark test case (illustrated in Fig. \ref{figA8}), against the smart agents equipped (via reinforcement learning) with the optimal behavioral policy ($\pi^*$).
Although efficacy of the implemented active control approach suffers from nonlinearly varying hydrodynamic loads (due to the presence of long-ranged flow-mediated interactions), the controlled agents are able to correct the induced deviations and eventually form the desired cloak (see Fig. \ref{figA8}-a).

It is also worth noting that in a model predictive control strategy, swimming actions (i.e. the control inputs) at each state, are determined by solving an on-line optimization problem.
Therefore, the main drawback of such an alternative approach is the potentially exorbitant on-line computational requirement \cite{ernst2008reinforcement}. 
Instead in the proposed reinforcement learning algorithm, the cloaking agents experientially learn optimal action policies through an \textit{off-line} trial-and-error training process (see e.g. Fig. \ref{figM3}). 
The learned behavioral policy (encoded in the so-called Q-matrix) then serves as the agents' decision-making intelligence, and the control actions are implemented on-line in the form of a simple table look-up (see section \ref{sec.Learning} for details).
To give an example, for the model predictive control implemented on the simple test case presented in Fig. \ref{figA8}, the on-line computation in taking each single action required $\sim 4.47 \times 10^{-1}$ s on a personal computer, whereas taking the optimal action based on the learned policy (i.e. through the table look-up) requires only $\sim 6.76 \times 10^{-5}$ s on the same machine (2019 Lenovo ThinkPad T470p; Intel Core i7-7700 HQ CPU @ 2.80 GHz; 8.00 GB RAM).
It is, however, to be mentioned that in principle, using the exact same state-action space as for the learning strategy, one can also pre-compute the optimal actions (as a function of initial state) in the predictive control scheme to make a look-up table \cite{bemporad2002explicit}.

%%%%%%%%%%%%%%%%%%%%%%%%%%%%%%%%%%%%%%%%%%%%%%%%%%%%%%%%%%%%%%%%%%%%%%%%%%%%%%%%%%%%%%%%%%%%%%%%%%%%%%%%%%%%

\subsection*{A.4. Incorporation of the Swimmers' Realistic Geometry}\hfill
\vspace{3mm}

In this section, we further explore effectiveness of the presented approach for cases where realistic geometry of the micro-swimmers is simulated in a three-dimensional (3D) setting.
Common approaches to provide precise description of swimmers' hydrodynamics in such settings include the boundary element method \cite[e.g.][]{liu2013helical,pimponi2016hydrodynamics} and the regularized Stokeslets approach \cite[e.g.][]{cortez2005method,walker2019filament}.
Formulation of the boundary element method (BEM) for Stokes flows is based on integrating a distribution of Stokeslets over the surface of each swimmer (see e.g. \cite{pozrikidis1992boundary,pozrikidis2002practical}).
As a result, the computational costs associated with this mesh-based method is relatively high, and can drastically increase with grid refinement \cite{liu2013helical}.
The regularized stokeslet method, on the other hand, can be a mesh-free approach for discretizing the boundary integral equations. A nearest-neighbor-based discretization of this approach (recently proposed by \cite{smith2018nearest}, has been shown to considerably reduce the computational cost without compromising the accuracy. Therefore, here we use this technique (as outlined by \cite{gallagher2018meshfree} to discretize the regularized Stokeslet boundary integral equations, and accurately model 3D geometry of the intruder as well as our swimming agents.

\begin{figure}
	\centering 
	\includegraphics[width=0.95\textwidth]{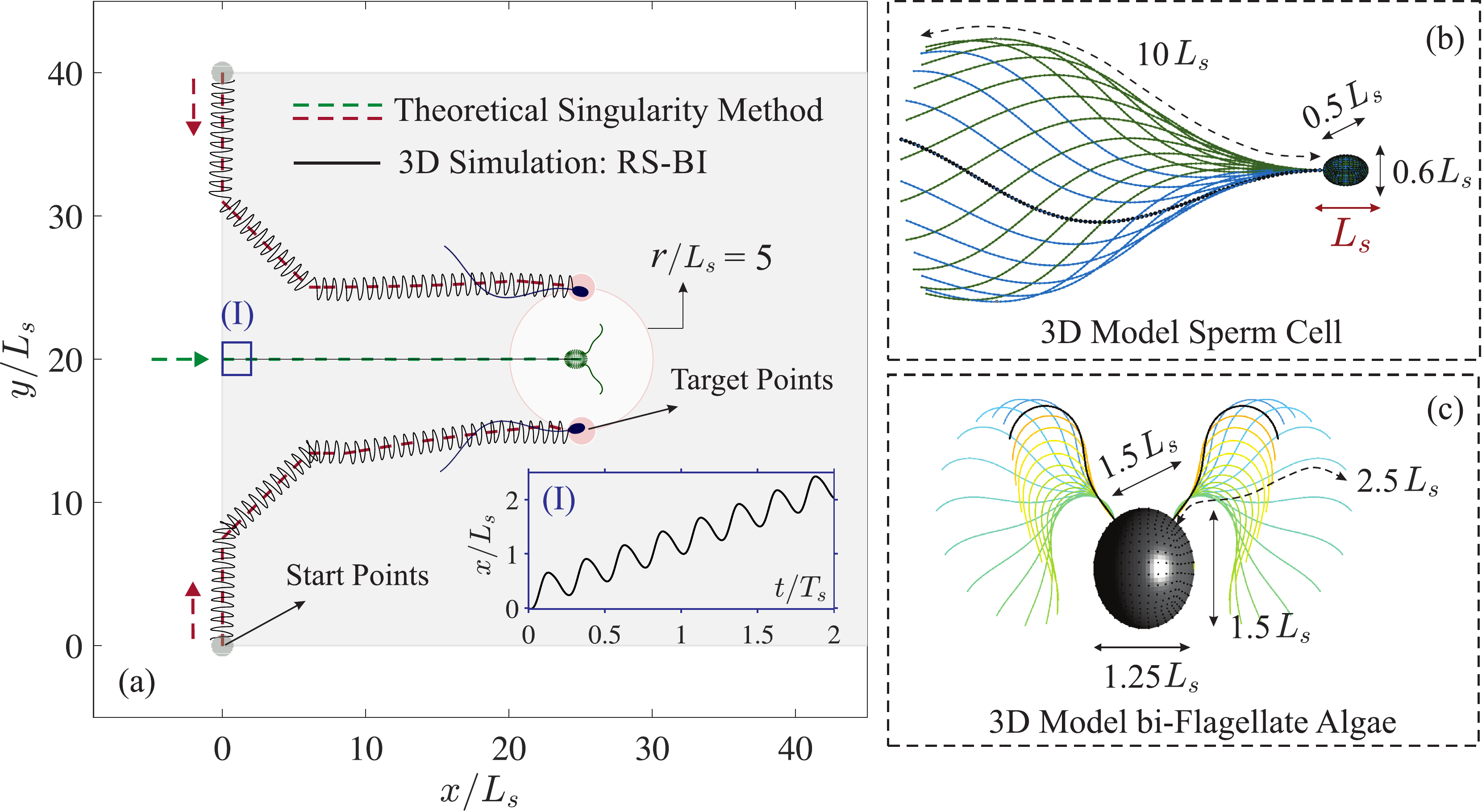}
	\caption { (a) Applicability of the learned behavioral policy for agent-control, when the simulation of hydrodynamics incorporates the swimmers' realistic 3D geometry. Specifically, the swimmers' hydrodynamics and their flow-mediated interactions are modeled using a high-accuracy 3D simulation framework (proposed by Ref. \cite{gallagher2018meshfree}) based on discretization of the regularized Stokeslets in boundary integral (RS-BI) formulation (solid trajectories). For the sake of comparison, we also plot (with dashed line) the trajectories presented in Fig. \ref{figM4}(f) associated with the test case conducted using ideal cloaking agents, modeled through the theoretical singularity method (as described in section \ref{sec.Cloaking}). In both cases, the cloaking agents are equipped with the optimal behavioral policy ($\pi^*$), obtained through the consecutive learning episodes demonstrated in Fig. \ref{figM3}. Here, the desired arrangement for the deployed agents is to symmetrically position themselves on the sides of intruder (with a separation distance $r_0/L_s =5$) that is properly achieved in both cases. The assigned target point for each agent is marked (in pink) at the final stages. Note that the RS-BI approach precisely models realistic geometry of the intruder and our swimming agents, as demonstrated in their final positions in panel (a). Here, as a benchmark, 3D models of self-propelled sperm cells (which propel based on the flexible-oar mechanism) are used to represent our pusher-type swimming agents. The 3D geometry of these model sperm cells and their beating pattern (i.e. swimming stroke) are presented in panel (b). The puller-type intruder, on the other hand, is simulated using a 3D model bi-flagellate algae, for which the swimming trajectory is also monitored over time, throughout the presented window in inset (I) of panel (a). The 3D geometry of the deployed bi-flagellate swimming cell along with its beating pattern is presented in panel (c).} 
	\label{figB2}
\end{figure}

Performance of the learned behavioral policy (as an agent-control strategy to realize desired cloaking arrangements around arbitrarily moving intruders) is then evaluated for cases where the swimmers' hydrodynamics is simulated using the described high-accuracy method (see Fig. \ref{figB2}).
In particular, we adapt the simulation framework and the open-source code provided by \cite{gallagher2018meshfree} to accurately model realistic 3D geometry of the intruder and cloaking agents.  
Here, for example, 3D models of self-propelled sperm cells (which propel based on the flexible-oar mechanism) are used to represent our pusher-type swimming agents. The puller-type intruder is also simulated using a 3D model bi-flagellate algae, in consistent with other sample tests presented throughout this article. Detailed (3D) geometry of the deployed model micro-swimmers and their beating patterns (i.e. swimming strokes) are also presented in panels (b)-(c) of Fig. \ref{figB2}.
Our numerical experiments further confirm that through implementing the sequence of actions provided by the learned policy ($\pi^*$), the swimming agents (simulated with realistic 3D geometries) can indeed form any desired cloaking arrangements (see e.g. Fig. \ref{figB2}) around an arbitrarily moving intruder (which is also simulated with its realistic 3D geometry).
In doing so, the swimming agents use an action space ($\mathcal{A}$) similar to the one described in section \ref{sec.Learning}.  Particularly, they can instantly turn to their right ($\theta \leftarrow \theta + \delta \theta$) or left ($\theta \leftarrow \theta - \delta \theta$) with different choices of angle $\delta \theta \in \lcb \pi/18,\pi/4,\pi/2 \rcb $. Note that such actions are chosen (with frequency $\tau_r^{-1}$) by the agents, based on the optimal behavioral policy ($\pi^*$) obtained through consecutive learning episodes demonstrated in Fig. \ref{figM3}.

\begin{figure}
	\centering 
	\includegraphics[width=0.8\textwidth]{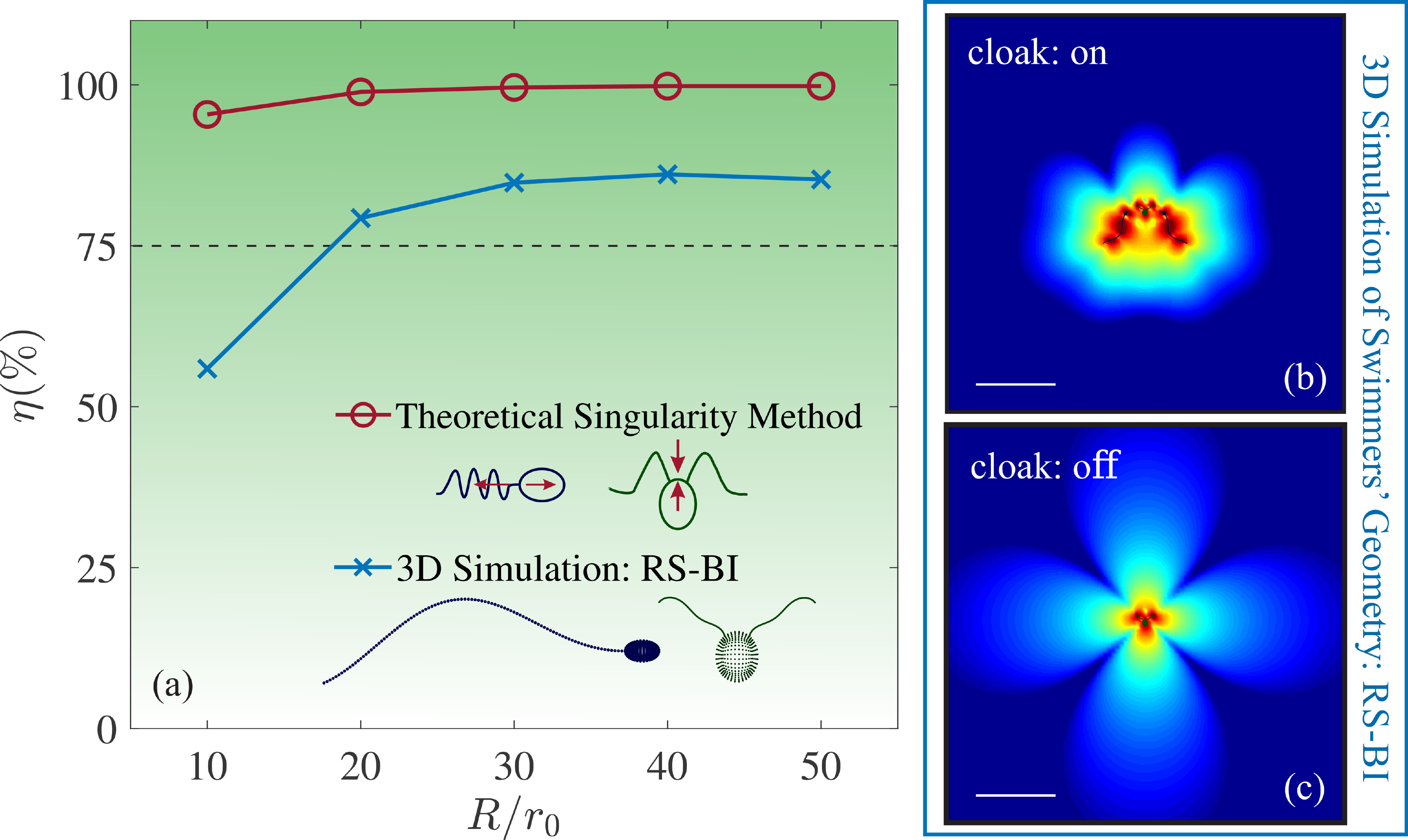}
	\caption {Effectiveness of the double-agent symmetric cloak formed at final stage of the numerical experiment (presented in Fig. \ref{figB2}) that incorporates the swimmers' realistic geometry by simulating hydrodynamics using a high-accuracy 3D framework based on discretization of the regularized Stokeslets in boundary integral (RS-BI) formulation. In panel (a) we compare the efficiency ($\eta$, quantified as a function of $R/r_0$) of the realized cloak (shown in red with `o' markers) to that associated with an ideal set of cloaking agents (shown in blue with `x' markers) modeled through the theoretical singularity method as described in section \ref{sec.Cloaking}. Inset also compares the model force-dipoles used in the singularity method, against the 3D realistic geometries used in the RS-BI simulation. In panel (b), we visualize performance of the cloak formed using the latter approach, where a realistic geometry of the intruder and cloaking agents are incorporated into the simulation of hydrodynamics. The color shading represents the magnitude of induced disturbing flows. For the sake of comparison, we also visualize (via color shading) the magnitude of fluid disturbances induced by an isolated intruder (c), when no cloaking agents are deployed (i.e. cloak: off). The presented snapshots correspond to the final stage of numerical experiment demonstrated in Fig. \ref{figB2}, and the scale bars denote $20L_s$.}
	\label{figB3}
\end{figure}

Effectiveness of the implemented cloak is then evaluated (see Fig. \ref{figB3}) for the benchmark numerical experiment presented in Fig. \ref{figB2}.
For the realized non-invasive cloak (in which the hydrodynamics are simulated by incorporating the swimmers' realistic geometry), our results reveal an efficiency ($\eta$) that reaches beyond 80\% and 85\% at ranges $R/r_0$ $\geq 20$ and $30$, respectively (Fig. \ref{figB3}).
This is owing to the fact that (as discussed in section \ref{sec.Interactions}) far-field of the disturbing flow induced by a micro-swimmer, can be well-described by the flow of a force dipole.
The difference (observed in Fig. \ref{figB3}) between far-field cloaking efficiencies of the presented test case (conducted using the RS-BI method) compared to those predicted in section \ref{sec.Cloaking} (using the theoretical singularity method) is due to the following fact.
In computing the cloaking efficiencies presented in section \ref{sec.Cloaking}, we use a set of artificial agents (i.e. swimming micro-robots) carefully designed to provide an ideally canceling set of dipole strengths for the specific intruder of interest (that is the $+\mathscr{D}/2$ dipole strength for each of the pusher-type cloaking agents, where $-\mathscr{D}$ is the dipole strength of the puller-type intruder). Therefore, the results associated with the theoretical singularity method, portray the theoretical limit for efficiencies that can be achieved using \textit{optimally designed} cloaking agents.  
Whereas, here we implement the RS-BI method to demonstrate an active cloaking test case using a set of micro-swimmers with geometric characteristics of swimming micro-organisms. In fact, 3D models of self-propelled sperm cells (with their realistic geometry as opposed to being carefully designed) are used to represent our pusher-type swimming agents, and a 3D model of bi-flagellate algae (with its realistic geometry) is used to represent the puller-type intruder (Fig. 12).
As a result, the associated dipole strengths (for the latter case) do not precisely cancel out, and thus the cloaking efficiency is expected to be less than those predicted (in section \ref{sec.Cloaking}) for an ideal set of cloaking agents (see the comparison in Fig. \ref{figB3}).
Nevertheless, in principle, by changing details of the swimmers' geometry (and/or swimming stroke), one can tune the force-dipoles representing far-field of their induced disturbances. 
Therefore, when using the sperm-like swimming micro-robots to cloak a specific intruder of interest, further optimization of their shape (and swimming stroke) can provide a perfectly canceling set of dipole strengths. This, however, is beyond the scope of this article and deserves an independent investigation.

In the end, we note that the implemented singularity method is based on modeling the disturbing flow induced by each micro-swimmer as the flow of a force dipole with \textit{fixed} strength/direction.
As discussed in section \ref{sec.Interactions}, this simple model has been widely used in the literature \cite{lauga2009hydrodynamics}, and its validity is particularly confirmed (through experimental measurements) for modeling the flow induced by E. coli bacteria \cite{drescher2011fluid}, as well as the time-averaged flow induced by C. reinhardtii algae \cite{drescher2010direct}.
However, it is worth mentioning that swimmers may induce time-varying flows characterized by a time-dependent dipole strength. 
Examples include the swimming cells simulated in the presented numerical experiment (Fig. \ref{figB2}) -- i.e. the bi-flagellate algae and spermatozoa, which transition between pusher- and puller-type dipoles over the course of their beat period.
In this section, we demonstrated (Figs. \ref{figB2}-\ref{figB3}) the possibility of using such swimming agents (with time-dependent flow fields) to form a desired arrangement and cloak an arbitrarily moving intruder (with a different time-varying disturbing flow).
To optimally conceal an intruder (over time) in such settings, however, the cloaking agents must additionally learn how to adjust their swimming gates (and/or beating patterns) in response to the time-varying flow induced by the intruder.
This, indeed, can be an interesting topic for future work.

On another note, in modeling swimmers with a natural counter-spinning of their cell body and flagella (such as E. coli bacteria), one can provide \cite[see e.g.][]{spagnolie2012hydrodynamics} a more accurate description of the near-field by supplementing the force-dipole model (decaying as 1/$r^2$) with a rotlet-dipole term (vanishing as 1/$r^3$).
Whilst micro-swimmers such as E. coli can swim along straight lines, the introduced rotlet-dipoles will not in general preserve the planarity of nearby swimmers. Thus, considerable care is required in devising 3D agent-control strategies for such systems.

\end{document}